\documentclass[12pt,thmsa]{article}
\usepackage{amsmath}
\usepackage[mathscr]{eucal}

\begin{document}

\author{
A.V. Shebeko$^{1 \dag}$ and P.A. Frolov$^{2}$
}
\title{A possible way for constructing generators of the Poincar\'{e} group in
quantum field theory}
\date{July 28 2011 }
\maketitle
\begin{center}
{\small\it $^{1}$ Institute for Theoretical Physics\\
National Research Center \lq\lq Kharkov Institute of Physics \& Technology\rq\rq\\
1 Akademicheskaya St., UA-61108 Kharkov 61108, Ukraine \\

\vskip 3mm

$^{2}$ Institute of Electrophysics \& Radiation
Technologies, NAS of Ukraine, \\
28 Chernyshevsky St., P.O. Box 8812, UA-61002 Kharkiv 61002,
Ukraine \\
$\dag$ {\it E-mail: shebeko@kipt.kharkov.ua }
}
\end{center}

\begin{tabular}{rr}
\hspace{80mm}                & To the memory of Mikhail Shirokov,\\
                             & the excellent scientist \\
                             & and modest person
\end{tabular}

\begin{abstract}
Starting from the instant form of relativistic quantum dynamics
for a system of interacting fields, where amongst the ten
generators of the Poincar\'{e} group only the Hamiltonian and the
boost operators carry interactions, we offer an algebraic method
to satisfy the Poincar\'{e} commutators. We do not need to employ
the Lagrangian formalism for local fields with the Noether
representation of the generators. Our approach is based on an
opportunity to separate in the primary interaction density a part
which is the Lorentz scalar. It makes possible apply the recursive
relations obtained in this work to construct the boosts in case of
both local field models (for instance with derivative couplings
and spins $\geq1$) and their nonlocal extensions. Such models are
typical of the meson theory of nuclear forces, where one has to
take into account vector meson exchanges and introduce
meson-nucleon vertices with cutoffs in momentum space.
Considerable attention is paid to finding analytic expressions for
the generators in the clothed-particle representation, in which
the so-called bad terms are simultaneously removed from the
Hamiltonian and the boosts. Moreover, the mass renormalization
terms introduced in the Hamiltonian at the very beginning  turn
out to be related to certain covariant integrals that are
convergent in the field models with appropriate cutoff factors.

\end{abstract}

\section{Introduction}
After Dirac \cite{Dirac49}, any relativistic quantum theory may be
so defined that the generator of time translations (Hamiltonian),
the generators of space translations (linear momentum), space
rotations (angular momentum) and Lorentz transformations (boost
operator) satisfy the well-known commutations. Basic ideas, put
forward by Dirac with his "front", "instant" and "point" forms of
the relativistic dynamics, have been realized in many relativistic
quantum mechanical models. In this context, the survey
\cite{KeisterPolyzou}, being remarkable introduction to a subfield
called the relativistic Hamiltonian dynamics, reflects various
aspects and achievements of relativistic direct interaction
theories. Among the vast literature on this subject we would like
to note an exhaustive exposition in lectures \cite{Bakker,Heinzl}
of the appealing features of the relativistic Hamiltonian dynamics
with an emphasis on "light-cone quantization". Following a
pioneering work \cite{Foldy61}, the term "direct" is related to a
system with a fixed number of interacting particles, where
interactions are rather direct than mediated through a field. In
the approach it is customary to consider such interactions
expressed in terms of the particle coordinates, momenta and spins.

This notion supplemented by the principle of cluster separability
(decomposition) was developed (see \cite{KeisterPolyzou} and refs.
therein) and applied to build up the so-called separable
interactions and relativistic center-of-mass variables for
composite systems \cite{Foldy61,KrajcikFoldy}. There were assumed
that the generators of the Poincar\'{e} group ($\Pi $) can be
represented as expansions on powers of $1/c^{2}$ or, more exactly,
$(v/c)^{2}$, where $v$ is a typical nuclear velocity (cf. the
$(p/m)$ expansion, introduced in \cite{Friar75,Friar77} in which
$m$ is the nucleon mass and $p$ is a typical nucleon momentum).
Afterwards, similar expansions were rederived and reexamined (with
new physical inputs) in the framework of another approach
\cite{GloeMuel81} (sometimes called the
Okubo-Gl\"{o}ckle-M\"{u}ller method \cite{Fuda}). There, starting
from a model Lagrangian for "scalar nucleons" interacting with a
scalar meson field (cf. the Wentzel model \cite{Wentzel}) the
authors showed (to our knowledge first) how the Hamiltonian and
the boost generator (the noncommuting operators), determined in a
standard manner \cite{Schwinger}, can be blockdiagonalized by one
and the same unitary transformation (UT) after Okubo \cite{Okubo}.
The corresponding blocks derived in leading order in the coupling
constant act in the subspace with a fixed nucleon number (the
nucleon "sector" of the full Fock space $R_{F}$).

In general, the work \cite{GloeMuel81} and its continuation
\cite{KruegGloe99,KruegGloe99C60} exemplify applications of local
relativistic quantum field theory (QFT), where the generators of
interest, being compatible with the basic commutation rules for
fields, are constructed within the Lagrangian formalism using the
N\"{o}ther theorem and its consequences. Although the available
covariant perturbation theory and functional-integral methods are
very successful when describing various relativistic and quantum
effects in the world of elementary particles, the Hamilton method
can be helpful too.

It is well known that it is the case, where one has to study
properties of strongly interacting particles, e.g., as in nuclear
physics with its problems of bound states for meson-nucleon
systems. Of course, any Hamiltonian formulation of field theory,
not being manifestly covariant, cannot be \textit{ab initio}
accepted as equivalent to the way after Feynman, Schwinger and
Tomonaga. However, in order to overcome the obstacle starting from
a field Hamiltonian $H$ one can consider it as one of the ten
infinitesimal operators (generators) of space-time translations
and pure Lorentz transformations that act in a proper Hilbert
space. Taken together they form a basis of the Poincar\'{e}-Lie
algebra with the aforementioned commutation relations to ensure
relativistic invariance (RI) in the Dirac sense, being referred to
the RI  as a whole \footnote{These relations will be recalled
below to fix the notations and simplify the reference processing}.
Our main purpose is to meet the Poincar\'{e} commutators for a
given interaction density which has the property to be a Lorentz
scalar in the Dirac (D) picture. Such a possibility may be
realized both in local and nonlocal models taking account their
invariance with respect to space translations. It turns out that
an algebraic method, which has been elaborated by us to get a
recursive solution of the problem in question, works also in
models (for instance, with derivative couplings and spin $\geq$1)
where only some part of the interaction density in the D picture
is the scalar.

As an illustration of our method, we will show its application for
a nonlocal extension of the Wentzel model. At the point, let us
remind of the nonlocal convergent field theory
\cite{ArnousHeitler,Guillot}, where a conventional interaction
Hamiltonian in the D picture (e.g., in quantum electrodynamics) is
replaced by a nonlocal interaction with a formfactor (FF) to
separate the field operators related to different points of the
Minkowski space (cf. monograph \cite{Efimov} and refs. therein in
which the same idea has been used directly for the initial action
integral). Unlike this in what follows, where we are addressing
the particle representation (see, for example, Chapter II in
lectures \cite{Friedrichs53} and Chapter IV of monograph
\cite{WeinbergBook1995}), the field concept has no its paramount
importance, being only a departure point for an alternative
consideration of the RI with particle creation. In the framework
of the particle representation a nonlocal Hamiltonian for
interacting particles can be built up by introducing some "cutoff"
function (shortly $g$-factor) in every vertex which is associated
with any particle creation and/or annihilation process. Such
cutoffs in momentum space may be done either phenomenologically or
with the aid of deeper physical reasonings as in case of the
meson-nucleon vertices that can be calculated in different quark
models (see, e.g., \cite{Plessas}).

As usually, the $g$-factors are needed, first of all, to carry out
finite intermediate calculations trying to remove ultraviolate
divergences inherent in local field models. One should emphasize
that we include them in the Yukawa-type interactions in the "bare"
particle representation (BPR) to derive or rather substantiate the
corresponding regularized interactions between the so-called
clothed particles (see Appendix C in \cite{FBS2010}). Their
falloff properties with the momenta increasing are also important
to do convergent calculations of strong and electromagnetic
 FFs (see Sec. 5).

Second, we will show how within the three-dimensional formalism
used here one define a covariant generating function for the
$g$-factors in case of a trilinear interaction. The function,
being dependent on some Lorentz scalars composed of the particle
three-momenta, plays a central role when integrating the
Poincar\'{e} commutators and obtaining the analytical
clothed-particle representation (CPR) expressions for the
Hamiltonian, the boost operators, the mass renormalization terms,
etc. \cite{SheShi00},\cite{KorShe04}.

Third, it is expected that by choosing appropriate $g$-factors (at
least, as square integrable functions of the particle momenta) one
can remove certain drawback of the initial local interaction not
to have a dense domain in $R_{F}$, i.e., not to be self-adjoint
and bounded below \footnote{The same is related to the boost
operators}. The delicate issue has been regarded in various papers
devoted to the Nelson model \cite{Nelson64} or "model with
persistent vacuum" (see, e.g., \cite{Eck70}, \cite{Albeverio73}
and refs. therein). It is true that the authors have confined
themselves to the explorations with a sharp cutoff and
antiparticles not included.

Along with a thorough option of the cutoffs for our nonlocal
extensions of the conventional field models with the threelinear
couplings the present research exemplifies one more realization of
a fruitful idea put forward in relativistic QFT by Greenberg and
Schweber \cite{GS58} and developed by other authors, in
particular, by Shirokov and his coworkers (see the survey
\cite{SheShi01} and refs. therein). First of all, we are keeping
in mind their notion of "clothed" particles which points out a
transparent way for including the so-called cloud or persistent
effects in a system of interacting fields (to be definite, mesons
and nucleons). It is achieved with the help of unitary clothing
transformations (UCTs) (see article \cite{KorCanShe}) that
implement the transition from the BPR to the CPR in the Hilbert
space $\mathcal{H}$ of meson-nucleon states. In the course of the
clothing procedure a large amount of virtual processes associated
with the meson absorption/emission, the $N\bar{N}$-pair
annihilation/production and other cloud effects turns out to be
accumulated in the creation (destruction) operators for the
clothed particles. The latter, being the quasiparticles of the
method of UCTs, must have the properties (charges, masses etc.) of
physical (observable) particles. Such a bootstrap reflects the
most significant distinction between the concepts of clothed and
bare particles.

As shown in \cite{SheShi01} the total Hamiltonian $H$ and the
three boost operators $\mathbf{N}=(N^{1},N^{2},N^{3})$ attain in
the CPR one and the same sparse structure in $\mathcal{H}$ due to
the elimination of the so-called bad terms that prevent the bare
vacuum (the state without "bare" particles \footnote{Following the
terminology accepted in \cite{SheShi01} (see also
\cite{KorCanShe}) every time when we say bare particles the latter
mean primary particles with physical masses}) and the bare
one-particle states to be the $H$ eigenstates. This result has
been obtained within a conventional local model of the PS
pion-nucleon coupling and, as we know, for local quantum theories
usually one goes from a relativistically invariant Lagrangian to
the corresponding Noether integrals that satisfy the Lie algebra
of the Poincar\'{e} group $\Pi $. Doing so, one can employ the
Belinfante ansatz to express $\mathbf{N}$ through the Hamiltonian
density (we recall it in Sec. 3). Here, trying to apply the UCT
method to nonlocal field models we will move in the opposite
direction, viz., from the fundamental Poincar\'{e} commutators
towards the RI as a whole.

However, before to apply the UCT method (in particular, beyond the
Lagrangian formalism with its local interaction densities) we will
show and compare the two algebraic procedures to solve the basic
commutator equations of $\Pi $ (see Sec. 2). One of them, proposed
here, has some touching points with the other developed in Refs.
\cite{Kita 66-72,Kita 68} and essentially repeated many years
after by Chandler \cite{Chandler}. In paper \cite{Kita 68} the
author considers three kinds of neutral spinless bosons and
nonlocal interaction between them in a relativistic version of the
Lee model with a cutoff in momentum space. A similar model for two
spinless particles has been utilized in \cite{Chandler} with a
Yukawa-type interaction that belongs to the realm of the so-called
models with persistent vacuum (see, for instance, \cite{Eck70}).

Certain resemblance between the present work and those
explorations is that we prefer to proceed, as previously
(\cite{SheShi00,KorShe04}), within a corpuscular picture (see
Chapter IV in monograph \cite{WeinbergBook1995}), where each of
the ten generators of the Poincar\'{e} group $\Pi $ (and not only
they) may be expressed
as a sum of products of particle creation and annihilation operators $%
a^{\dagger }(n)$ and $a(n)$ $(n=1,2,...)$ e.g., bosons and/or
fermions. Some mathematical aspects of the corpuscular notion were
formulated many years ago in \cite{Friedrichs} (Chapter III). As
in \cite{WeinbergBook1995}, a label $n$ is associated with all the
necessary quantum numbers for a single particle:
its momentum $\mathbf{p}_{\,n}$\footnote{%
Or the 4-momentum $p_{\,n}=(p_{\,n}^0,\mathbf{p}_{\,n})$ on the mass shell $%
p_{\,n}^2=p_{\,n}^{02}-\mathbf{p}_{\,n}^2=m_n^2$ with the particle
mass $m_n $}, spin $z$-component (or for massless particles,
helicity) $\mu _{\,n}$, and species $\xi _{\,n}$. The operators
$a^{\dagger }(n)$ and $a(n)$ satisfy the standard (canonical)
commutation relations such as Eqs. (4.2.5)-(4.2.7) in
\cite{WeinbergBook1995}.

In the framework of such a picture the Hamiltonian of a system of
interacting mesons and nucleons can be written as
\begin{equation}
H=\sum_{C=0}^\infty \sum_{A=0}^\infty H_{CA},  \label{1.1}
\end{equation}
\begin{eqnarray}
H_{CA} = \int \!\!\!\!\!\!\!\sum H_{CA}(1^{\prime },2^{\prime },
...,n_{C}';1\mathbf{,}2\mathbf{,}\mathbf{...,}n_{A})
\nonumber \\
\times a^{\dag}(1^{\prime })a^{\dag}(2^{\prime })
...a^{\dag}(n_{C}')a(n_{A})...a(2)a(1), \label{1.2}
\end{eqnarray}
where the capital $C(A)$  denotes the particle-creation
(annihilation) number for the operator substructure $H_{CA}$.
Sometimes we say that the latter belongs to the class
$[C\mathbf{.}A]$ (cf. the terminology from \cite{SheShi01}).
Operation $\int\!\!\!\!\! \sum $ implies all necessary summations
over discrete indices and covariant integrations over continuous
spectra.

Further, it is proved \cite{WeinbergBook1995} that the $S$-matrix
meets the so-called cluster decomposition principle (see, e.g.,
\cite{WichCrich}) if the coefficient functions $H_{CA}$ embody a
single three-dimensional momentum-conservation delta function,
viz.,
\[
H_{CA}(1^{\prime },2^{\prime },...,C;1\mathbf{,}2\mathbf{,...,}A)=\delta (%
\mathbf{p}_1^{\prime }+\mathbf{p}_2^{\prime }+...+\mathbf{p}_C^{\prime }-%
\mathbf{p}_1-\mathbf{p}_2-...-\mathbf{p}_A)
\]
\begin{equation}
\times h_{CA}(p_1^{\prime }\mu _1^{\prime }\xi _1^{\prime
},p_2^{\prime }\mu _2^{\prime }\xi _2^{\prime },...,p_C^{\prime
}\mu _C^{\prime }\xi _C^{\prime };p_1\mu _1\xi _1,p_2\mu _2\xi
_2,...,p_A\mu _A\xi _A),  \label{1.3}
\end{equation}
where the $c$-number coefficients $h_{CA}$ do not contain delta
function.

Following the guideline ``to free ourselves from any dependence on
pre-existing field theories ''(cit. from \cite{WeinbergBook1995}
on p.175), the three boost operators $\mathbf{N=}(N^1,N^2,N^3)$
can be written as
\begin{equation}
\mathbf{N}=\sum_{C=0}^\infty \sum_{A=0}^\infty \mathbf{N}_{CA},
\label{1.4}
\end{equation}
\[
\ \mathbf{N}_{CA} =\int \!\!\!\!\!\!\!\sum
\mathbf{N}_{CA}(1^{\prime },2^{\prime
},...,n_{C}';1\mathbf{,}2\mathbf{,...,}n_{A})
\]
\begin{equation}
\times a^{\dagger }(1^{\prime })a^{\dagger }(2^{\prime })%
...a^{\dagger }(n_{C}')a(n_{A})...a(2)a(1). \label{1.5}
\end{equation}

One of our purposes is to find some links between the coefficients
in the r.h.s. of Eqs. (\ref{1.2}) and (\ref{1.5}), compatible with
the fundamental relations of the Lie algebra for $\Pi $, that are
given for convenience in Sec. 2.

In turn, the operator $H$, being divided into the no-interaction
part $H_{F}$ and the interaction $H_{I}$, owing to its
translational invariance allows $H_{I}$ to be written as
\begin{equation}
H_{I}=\int H_{I}(\mathbf{x})d\mathbf{x}. \label{1.6}
\end{equation}

Our consideration is focused upon various field models (local and
nonlocal) in which the interaction density $H_{I}(\mathbf{x})$
consists of scalar $H_{sc}(\mathbf{x})$ and nonscalar
$H_{nsc}(\mathbf{x})$ contributions,
\begin{equation}
H_{I}(\mathbf{x})= H_{sc}(\mathbf{x})+H_{nsc}(\mathbf{x}),
 \label{1.7}
\end{equation}
where the property to be a scalar means
\begin{equation}
U_{F}(\Lambda)H_{sc}(x)U_{F}^{-1}=H_{sc}(\Lambda
x),\,\,\,\,\forall x=(t,\mathbf{x})
 \label{1.8}
\end{equation}
for all Lorentz transformations $\Lambda$. Henceforth, for any
operator $O(\mathbf{x})$ in the Schr\"{o}dinger (S) picture it is
introduced its counterpart
\[
O(x)=e^{iH_{F}t}O(\mathbf{x})e^{-iH_{F}t}
\]
in the Dirac (D) picture.

In this context we would like to remind that in "...theories with
derivative couplings or spins $j\geq1$, it is not enough to take
Hamiltonian as the integral over space of a scalar interaction
density; we also need to add non-scalar terms to the interaction
density to compensate non-covariant terms in the propagators"
(quoted from Chapter 7 in \cite{WeinbergBook1995}). Such a
situation has been considered recently for interacting vector
mesons and nucleons in the field-theoretical treatment
\cite{Bonn09,FBS2010} of nucleon-nucleon scattering. In any case,
as will be shown, the existence of division (\ref{1.7}) makes it
possible to use and extend the available experience
\cite{SheShi01} in constructing the boost generators for a given
$H_{I}(\mathbf{x})$.

As previously \cite{KorShe04,SheShi01}, special attention in our
work is paid to the inclusion in $H$ finite "mass-renormalization"
terms that play an important role in ensuring the RI \cite{Kita
68}. We stress "finite" since in what follows in order to get rid
of the well known difficulties with divergences certain emphasis
is made on nonlocal field models with a covariant cutoff. Thereby
we prefer to deal with introducing cutoff functions in momentum
space that is convenient for calculations of the S matrix (cf. a
relativistic nonlocal field model proposed in \cite{Shir2002} with
cutoffs in coordinate space).

After this introduction we arrive to Sec. 2 which is devoted to
some preliminaries concerning the underlying problem. In Sec. 3,
by considering nonperturbative and perturbative recipes for
ensuring the RI, we recall a number of relevant definitions from
local QFT. Such a reminder enables us to set bridges between a
traditional approach in QFT and direct algebraic means proposed
here. By uniting our algebraic approach with the notion of clothed
particles in QFT, in Sec. 4 we are seeking the boost operators in
the CPR. Along the headline we introduce a nonstandard definition
of the so-called mass renormalization terms and show their
importance for ensuring the RI within a wide class of field models
(local and nonlocal). Sec. 5 is contained explicit expressions for
the interactions (quasipotentials) between the spinless scalar and
charged bosons and the corresponding renormalization integrals.

The Appendices A, B and C are contained, respectively,

a) formulae for the Poincar\'{e} generators of free pions and
nucleons in the corpuscular picture

b) equal-time commutators for the pion-nucleon interaction
densities with a nonlocal trilinear coupling

c)evaluation of an integral that determines the mass
renormalization term in case of a relativistic nonlocal model for
interacting spinless neutral and charged bosons.

\section{Basic equations in relativistic theory with particle creation and
annihilation} For convenience, the Poincar\'{e} generators can be
divided into the three kinds for:
\newline no-interaction generators
\begin{equation}
\lbrack P_i\mathbf{,}P_j]=0,\,\,\,\,\,\,\,\,\,[J_i\mathbf{,}%
J_j]=i\varepsilon _{ijk}J_k,\,\,\,\,\,\,\,\,\,{\ }[J_i\mathbf{,}%
P_j]=i\varepsilon _{ijk}P_k,  \label{2.9}
\end{equation}
generators linear in $H$ and $\mathbf{N}$
\begin{equation}
\lbrack \mathbf{P,}H]=0,\,\,\,\,\,\,\,\,\,[\mathbf{J,}H]=0,\,\,\,\,\,\,\,\,%
\,[J_i\mathbf{,}N_j]=i\varepsilon _{ijk}N_k,\,\,\,\,\,\,\,\,\,[P_i\mathbf{,}%
N_j]=i\delta _{ij}H,  \label{2.10}
\end{equation}
and ones nonlinear in $H$ and $\mathbf{N}$
\begin{equation}
\lbrack H,\mathbf{N}]=i\mathbf{P},\,\,\,\,\,\,\,\,\,[N_i\mathbf{,}%
N_j]=-i\varepsilon _{ijk}J_k,  \label{2.11}
\end{equation}
\[
(i,j,k=1,2,3),
\]
where $\mathbf{P}=(P^1,P^2,P^3)$ and $\mathbf{J}=(J^1,J^2,J^3)$
are the linear momentum and angular momentum operators,
respectively. In this context, let us remind that in the instant
form of relativistic dynamics after Dirac \cite {Dirac49} only the
Hamiltonian and the boost operators carry interactions with
conventional partitions
\begin{equation}
H=H_F+H_I  \label{2.12}
\end{equation}
and
\begin{equation}
\mathbf{N}=\mathbf{N}_F+\mathbf{N}_I,  \label{2.13}
\end{equation}
while $\mathbf{P=P}_F$ and $\mathbf{J=J}_F$. In short notations,
we distinguish the set $G_F=\{H_F,\mathbf{P}_F,\mathbf{J}_F,%
\mathbf{N}_F\}$ for free particles and the set $G=\{H,\mathbf{P}_F,\mathbf{J}%
_F,\mathbf{N}\}$ for interacting particles.

In turn, every operator $H_{CA}$ can be represented as
\begin{equation}
H_{CA}=\int H_{CA}(\mathbf{x})d\mathbf{x,}  \label{2.14}
\end{equation}
if one uses the formula
\[
\delta (\mathbf{p-p}^{\prime })=\frac 1{(2\pi )^3}\int e^{i(\mathbf{p-p}%
^{\prime })\mathbf{x}}d\mathbf{x.}
\]
Thus, we come to the form well known from local field models,
\begin{equation}
H=\int H(\mathbf{x})d\mathbf{x}  \label{2.15}
\end{equation}
with the density
\begin{equation}
H(\mathbf{x})=\sum_{C=0}^\infty \sum_{A=0}^\infty
H_{CA}(\mathbf{x}). \label{2.16}
\end{equation}
For instance, in case with $C=A=2,$
\begin{equation}
H_{22}(1^{\prime },2^{\prime };1,2)=\delta (\mathbf{p}_1^{\prime }+\mathbf{p}%
_2^{\prime }-\mathbf{p}_1-\mathbf{p}_2)h(1^{\prime },2^{\prime
};1,2) \label{2.17}
\end{equation}
and
\begin{eqnarray}
H_{22}(\mathbf{x}) =\frac 1{(2\pi )^3}\int \!\!\!\!\!\!\!\sum \exp
[-i(\mathbf{p}_1^{\prime }+\mathbf{p}_2^{\prime }-\mathbf{p}_1-\mathbf{p}_2)%
\mathbf{x]}  \nonumber \\
\times h(1^{\prime },2^{\prime };1,2)a^{\dagger }\left( 1^{\prime
}\right) a^{\dagger }\left( 2^{\prime }\right) a\left( 2\right)
a\left( 1\right) . \label{2.18}
\end{eqnarray}

Further, we will employ the transformation properties of the
creation and annihilation operators with respect to $\Pi $. For
example, in case of a massive particle with the mass $m$ and spin
$j$ one considers that
\begin{equation}
U_F(\Lambda ,b)a^{\dagger }(p,\mu )U_F^{-1}(\Lambda
,b)=e^{i\Lambda pb}D_{\mu ^{\prime }\mu }^{(j)}(W(\Lambda
,p))a^{\dagger }(\Lambda p,\mu ^{\prime }),  \label{2.19}
\end{equation}
\[
\forall \Lambda \in L_{+}\,\,\mathrm{{\
and\,\,arbitrary\,\,spacetime\,\,shifts}\,\,b=(b^0,\mathbf{b})}
\]
with $D$-function whose argument is the Wigner rotation $W(\Lambda ,p),$ $%
L_{+}$ the homogeneous (proper) orthochronous Lorentz group. The
correspondence $(\Lambda ,b)\rightarrow U_F(\Lambda ,b)$ between elements $%
(\Lambda ,b)$ $\in $ $\Pi $ and unitary transformations
$U_F(\Lambda ,b)$
realizes an irreducible representation of $\Pi $ on the Hilbert space $%
\mathcal{H}$ (to be definite) of meson-nucleon states. In this
context, it is convenient to employ the operators $a(p,\mu
)=a(\mathbf{p},\mu )\sqrt{p_0}$ that meet the covariant
commutation relations
\begin{eqnarray}
\lbrack a(p^{\prime },\mu ^{\prime }),a^{\dagger }(p,\mu )]_{\pm }
&=&p_0\delta (\mathbf{p-p}^{\prime })\delta _{\mu ^{\prime }\mu },
\nonumber
\\
\lbrack a(p^{\prime },\mu ^{\prime }),a(p,\mu )]_{\pm }
&=&[a^{\dagger }(p^{\prime },\mu ^{\prime }),a^{\dagger }(p,\mu
)]_{\pm }=0.  \label{2.20}
\end{eqnarray}
Here $p_0=\sqrt{\mathbf{p}^2+m^2}$ is the fourth component of the
4-momentum $p=(p_0,\mathbf{p}).$

The relativistic invariance (RI) implies
\begin{equation}
U_F(\Lambda ,b)H_{22}(x)U_F^{-1}(\Lambda ,b)=H_{22}(\Lambda
x+b),\,\,\,\,\forall x=(t,\mathbf{x}).  \label{2.21}
\end{equation}
Accordingly this definition we have
\begin{eqnarray}
H_{22}(x) &=&\frac 1{(2\pi )^3}\int \!\!\!\!\!\!\!\sum \exp
[i(p_1^{\prime }+p_2^{\prime }-p_1-p_2)x]  \nonumber \\
&&\times h(1^{\prime },2^{\prime };1,2)a^{\dagger }\left(
1^{\prime }\right) a^{\dagger }\left( 2^{\prime }\right) a\left(
2\right) a\left( 1\right) .  \label{2.22}
\end{eqnarray}
With the aid of Eq. (\ref{2.19}) it is easily seen that condition
(\ref {2.21}) imposes the following constraint upon the
$h$-coefficients in the r.h.s. of Eq. (\ref{2.22}):
\[
D_{\eta _1^{\prime }\mu _1^{\prime }}^{(j_1^{\prime })}(W(\Lambda
,p_1^{\prime }))D_{\eta _2^{\prime }\mu _2^{\prime
}}^{(j_2^{\prime })}(W(\Lambda ,p_2^{\prime }))D_{\eta _1\mu
_1}^{(j_1)*}(W(\Lambda ,p_1))D_{\eta _2\mu _2}^{(j_2)*}(W(\Lambda
,p_2))
\]
\begin{equation}
\times h(p_1^{\prime }\mu _1^{\prime },p_2^{\prime }\mu _2^{\prime
};p_1\mu _1,p_2\mu _2)=h(\Lambda p_1^{\prime }\eta _1^{\prime
},\Lambda p_2^{\prime }\eta _2^{\prime };\Lambda p_1\eta
_1,\Lambda p_2\eta _2).  \label{2.23}
\end{equation}
Of course, summations over all dummy labels are implied.

\section{Nonperturbative and perturbative recipes for ensuring relativistic
invariance} We will find an effective way to meet the commutation
relations of the Lie algebra for the Poincar\'{e} group in terms
of the creation (annihilation) operators of particles in momentum
space with the concept of fields not to be used. Our algebraic
approach is aimed at the ensuring of RI \textit{as a whole }unlike
the Lagrangian formalism, where requirements of relativistic
symmetry are \textit{manifestly }provided at the beginning.
Meanwhile, we strive to go out beyond the traditional QFT with
local Lagrangian densities via a special regularization of
interactions in a total initial Hamiltonian.

\subsection{Definitions of the Poincar\'{e} generators in a local QFT. Application to
interacting pion and nucleon fields} It is well known that within
the Lagrangian formalism the 4-vector $P^\mu =(H,\mathbf{P})$ is
determined by the Noether integrals
\begin{equation}
P^{\,\nu }=\int \mathcal{T}^{0\nu
}(\mathbf{x})d\mathbf{x}\,\,\,\,\,(\nu =0,1,2,3),  \label{3.24}
\end{equation}
where $\mathcal{T}^{0\nu }(\mathbf{x})$ are the components of the
energy-momentum tensor density $\mathcal{T}^{\mu \nu }(x)$ at
$t=0.$

Other Noether integrals are expressed through the angular-momentum
tensor density
\begin{equation}
\mathcal{M}^{\beta [\mu \nu ]}(x)=x^{\,\mu }\mathcal{T}^{\beta \nu
}(x)-x^{\,\nu }\mathcal{T}^{\beta \mu }(x)+\Sigma ^{\beta [\mu \nu
]}(x), \label{3.25}
\end{equation}
that contains, in general, so-called polarization part $\Sigma
^{\beta [\mu \nu ]}\footnote{
Henceforth, the symbol $[\alpha ,\beta ]$ for any labels $\alpha $ and $%
\beta $ means the property $f^{[\beta ,\alpha ]}=-f^{[\alpha
,\beta ]}$ for its carrier $f.$}$ associated with spin degrees of
freedom. Namely, the six independent integrals
\begin{equation}
M^{\mu \nu }=\left. \int \mathcal{M}^{0[\mu \nu
]}(x)d\mathbf{x}\right| _{t=0}  \label{3.26}
\end{equation}
are considered as the generators of space rotations
\begin{equation}
J^i=\varepsilon _{ikl}M^{kl}\,\,\,\,\,(i,k,l=1,2,3)  \label{3.27}
\end{equation}
and the boosts
\begin{equation}
N^k\equiv M^{0k}=-\int
x^k\mathcal{T}^{00}(\mathbf{x})d\mathbf{x+}\int \Sigma
^{0[0k]}(\mathbf{x})d\mathbf{x},\,(k=1,2,3).  \label{3.28}
\end{equation}

As an illustration, for interacting pion and nucleon fields with
the PS coupling starting from the Lagrangian density after
\cite{SchweberBook} (cf. model (13.42) in \cite{BjorkenDrell} with
its non-hermitian Lagrangian density)
\[
\mathcal{L}_{SCH}(x) =\frac 12\bar{\psi}_H(x)(i\gamma ^{\,\mu }
\overrightarrow{\partial }_{\,\mu }-m_0)\psi _H(x)+\frac
12\bar{\psi} _H(x)(-i\gamma ^{\,\mu }\overleftarrow{\partial
}_{\,\mu }-m_0)\psi _H(x)
\]
\begin{equation}
+\frac 12[\partial _{\,\mu }\varphi _H(x)\partial ^{\,\mu }\varphi
_H(x)-\mu _0^2\varphi _H^2(x)]-ig_0\bar{\psi}_H(x)\gamma _5\psi
_H(x)\varphi _H(x),  \label{3.29}
\end{equation}
one has (omitting argument $x$):

i) Euler-Lagrange equations

\begin{equation}
\frac{\partial \mathcal{L}_{SCH}}{\partial \bar{\psi}_H}-\partial _{\,\mu }%
\frac{\partial \mathcal{L}_{SCH}}{\partial \bar{\psi}_{H\,\mu }}%
=0,\,\,\,\,\,\,\frac{\partial \mathcal{L}_{SCH}}{\partial \psi
_H}-\partial _{\,\mu }\frac{\partial \mathcal{L}_{SCH}}{\partial
\psi _{H\,\mu }}=0, \label{3.30}
\end{equation}
or
\[
\frac 12(i\gamma ^{\,\mu }\overrightarrow{\partial }_{\,\mu
}-m_0)\psi _H=ig_0\gamma _5\psi _H\varphi _H,
\]
\begin{equation}
\frac 12\bar{\psi}_H(-i\gamma ^{\,\mu }\overleftarrow{\partial
}_{\,\mu }-m_0)=ig_0\bar{\psi}_H\gamma _5\varphi _H \label{3.31}
\end{equation}
with ''bare'' nucleon mass $m_0,$ pion mass $\mu _0$ and coupling constant $%
g_{0},$

ii) energy-momentum tensor density
\[
\mathcal{T}_{SCH}^{\mu \nu }=\frac{\partial \mathcal{L}_{SCH}}{\partial \bar{%
\psi}_{H\,\mu }}\bar{\psi}_H^{\,\nu }+\frac{\partial \mathcal{L}_{SCH}}{%
\partial \psi _{H\,\mu }}\psi _H^{\,\nu }+\frac{\partial \mathcal{L}_{SCH}}{%
\partial \varphi _{H\,\mu }}\varphi _H^{\,\nu }-g^{\mu \nu }\mathcal{L}%
_{SCH}
\]
\begin{equation}
\equiv \mathcal{T}_N^{\mu \nu }+\mathcal{T}_{\,\pi }^{\mu \nu }+%
\mathcal{T}_I^{\mu \nu },  \label{3.32}
\end{equation}
where
\begin{equation}
\mathcal{T}_N^{\mu \nu }=\frac i2\bar{\psi}_H\gamma ^{\,\mu
}\partial
^{\,\nu }\psi _H-\frac i2\gamma ^{\,\mu }\psi _H\partial ^\nu \bar{\psi}%
_H-g^{\mu \nu }\mathcal{L}_N,  \label{3.33}
\end{equation}
\begin{equation}
\mathcal{T}_{\,\pi }^{\,\mu \nu }=\partial ^{\,\mu }\varphi
_H\partial ^{\,\nu }\varphi _H-g^{\mu \nu }\mathcal{L}_{\,\pi },
\label{3.34}
\end{equation}
\begin{equation}
\mathcal{T}_I^{\,\mu \nu }=ig_0g^{\mu \nu }\bar{\psi}_H\gamma
_5\psi _H\varphi _H,  \label{3.35}
\end{equation}
iii) polarization contribution
\begin{equation}
\Sigma _{SCH}^{\,\beta [\mu \nu ]}=\frac 12i\bar{\psi}_H\{\gamma
^{\,\beta }\Sigma ^{\,\mu \nu }+\Sigma ^{\,\mu \nu }\gamma
^{\,\beta }\}\psi _H, \label{3.36}
\end{equation}
where
\[
\Sigma ^{\,\mu \nu }=\frac i4[\gamma ^{\,\mu },\gamma ^{\,\nu }].
\]
In formulae (\ref{3.29})-(\ref{3.35}) unlike operators $O(x)$ in
the D picture, we have operators
\[
O_H(x)=e^{iHt}O(\mathbf{x})e^{-iHt},
\]
in the Heisenberg (H) picture. As before we prefer to employ the
definitions:
\[
\{\gamma ^{\,\mu },\gamma ^{\,\nu }\}=2g^{\mu \nu },\gamma _{\,\mu
}^{\dagger }=\gamma _0\gamma _{\,\mu }\gamma _0,\{\gamma _{\,\mu
},\gamma _5\}=0,\gamma _5^{\dagger }=\gamma _0\gamma _5\gamma
_0=-\gamma _5.
\]

The corresponding Hamiltonian density is given by
\begin{equation}
H_{SCH}(\mathbf{x})=\mathcal{T}^{00}_{SCH}(\mathbf{x})=H_{ferm}^{0}(\mathbf{x})+H_{\pi}^{0}
(\mathbf{x})+V_{ps}^{0}(\mathbf{x}),  \label{3.37}
\end{equation}
where
\begin{equation}
H^{0}_{ferm}(\mathbf{x})=\frac 12\bar{\psi}(\mathbf{x})[-i\overrightarrow{\gamma }%
\overrightarrow{\partial }+m_0]\psi (\mathbf{x})+\frac 12\bar{\psi}(\mathbf{x%
})[+i\overleftarrow{\gamma }\overleftarrow{\partial }+m_0]\psi
(\mathbf{x}), \label{3.38}
\end{equation}
\begin{equation}
H^{0}_\pi (\mathbf{x})=\frac 12\left[ \pi ^2(\mathbf{x})+\nabla \varphi (\mathbf{%
x})\nabla \varphi (\mathbf{x})+\mu _0^2\varphi
^2(\mathbf{x})\right] , \label{3.39}
\end{equation}
\begin{equation}
V_{ps}^{0}(\mathbf{x})=-ig_0\bar{\psi}(\mathbf{x})\gamma _5\psi (\mathbf{x})\varphi (%
\mathbf{x}).  \label{3.40}
\end{equation}
Following a common recipe (see, e.g., Sec 7.5 in
\cite{WeinbergBook1995}) we have introduced the canonical
conjugate variable
\begin{equation}
\pi (\mathbf{x})\equiv \left. \dot{\varphi}(x)\right| _{t=0}
\label{3.41}
\end{equation}
for the pion field. One should note that the second integral in
the r.h.s.
of Eq. (\ref{3.28}) does not contribute to the model boost since operator (%
\ref{3.36}) is identically equal zero. In fact,
\[
\gamma ^0\Sigma ^{\,0k}+\Sigma ^{\,0k}\gamma ^0=\frac i4\{\gamma
^0[\gamma ^0\gamma ^k-\gamma ^k\gamma ^0]+[\gamma ^0\gamma
^k-\gamma ^k\gamma ^0]\gamma ^0\}
\]
\[
=\frac i4\{\gamma ^k-\gamma ^0\gamma ^k\gamma ^0+\gamma ^0\gamma
^k\gamma ^0-\gamma ^k\}=0.
\]
Thus we have
\begin{equation}
\mathbf{N}_{SCH}=-\int
\mathbf{x}\mathcal{T}_{SCH}^{00}(\mathbf{x})d\mathbf{x}=-\int
\mathbf{x}H_{SCH}(\mathbf{x})d\mathbf{x}. \label{3.42}
\end{equation}

The relation (\ref{3.42}) exemplifies the so-called Belinfante
ansatz:
\begin{equation}
\mathbf{N}=-\int\mathbf{x}H(\mathbf{x})d\mathbf{x},
 \label{3.43}
\end{equation}
which, as it has first been shown in \cite{Bel40}, holds for any
local field model with a symmetrized density tensor
$\mathcal{T}^{\mu \nu }(x)=\mathcal{T}^{\nu \mu }(x)$. Such a
representation helps \cite{SheShi01} to implement a simultaneous
blockdiagonalization of the Hamiltonian and the generators of
Lorentz boosts in the CPR \footnote{The relation (\ref{3.43}) also
has turned out to be useful when formulating a local analog of the
Siegert theorem in the covariant description of electromagnetic
interactions with nuclei \cite{She90}}. We shall come back to this
point later.

 By passing, we would like to note that the tensor (\ref{3.32})
being symmetrized after Belinfante can be written in the form
\begin{equation}
\mathcal{T}_{sym}^{\mu \nu }=\mathcal{T}_{N,sym}^{\mu \nu
}+\mathcal{T}_{\pi}^{\mu \nu }+\mathcal{T}_{I}^{\mu \nu },
 \label{3.44}
\end{equation}
\[
 \mathcal{T}_{N,sym}^{\mu \nu }=\frac{i}{4}(
\bar{\psi}_{H}(x)\gamma^{\mu}\partial^{\nu}\psi_{H}(x)+
\bar{\psi}_{H}(x)\gamma^{\nu}\partial^{\mu}\psi_{H}(x)
\]
\[
-\partial^{\nu}\bar{\psi}_{H}(x)\gamma^{\mu}\psi_{H}(x)-
\partial^{\mu}\bar{\psi}_{H}(x)\gamma^{\nu}\psi_{H}(x))
 -g^{\mu\nu}\mathcal{L}_{N}.
\]

Further, the Hamiltonian density can be represented as
\begin{equation}
H_{SCH}(\mathbf{x})=H_{F}(\mathbf{x})+H_{I}(\mathbf{x})
\label{3.45}
\end{equation}
with the free part
\begin{equation}
H_{F}(\mathbf{x})=H_{\pi}(\mathbf{x})+H_{ferm}(\mathbf{x})
\label{3.46}
\end{equation}
and the interaction density
\begin{equation}
H_{I}(\mathbf{x})=V_{ps}(\mathbf{x})+H_{ren}(\mathbf{x}),\,\,
V_{ps}(\mathbf{x})=ig\bar{\psi}(\mathbf{x})\gamma_{5}\psi(\mathbf{x})\varphi(\mathbf{x}),
\label{3.47}
\end{equation}
where we have introduced the mass and vertex counterterms:
\begin{equation}
H_{ren}(\mathbf{x})=M_{ren}^{mes}(\mathbf{x})+M_{ren}^{ferm}(\mathbf{x}%
)+H_{ren}^{int}(\mathbf{x}),  \label{3.48}
\end{equation}
\[
\ M_{ren}^{mes}(\mathbf{x})=\frac 12(\mu
_0^2-\mu^{2}_{\pi})\varphi ^2(\mathbf{x}),
\]
\[
\ M_{ren}^{ferm}(\mathbf{x})=(m_0-m)\bar{\psi}(\mathbf{x})\psi
(\mathbf{x})
\]
and
\[
\ H_{ren}^{int}(\mathbf{x})=i(g_0-g)\bar{\psi}(\mathbf{x})\gamma _5\psi (%
\mathbf{x})\varphi (\mathbf{x}).
\]

One should note that the densities in Eqs.
(\ref{3.46})-(\ref{3.47}) are obtained from Eqs.
(\ref{3.38})-(\ref{3.39}) replacing the bare values $m_{0}$,
$\mu_{0}$ and $g_{0}$, respectively, by the "physical" values $m$,
$\mu_{\pi}$ and $g$. Such a transition can be done via the
mass-changing Bogoliubov-type unitary transformations (details see
in \cite{KorCanShe}). In particular, the fields involved can be
expressed through the set $\alpha={a^{\dag}(a), b^{\dag}(b),
d^{\dag}(d)}$ of the creation (destruction) operators for the bare
pions and nucleons with the physical masses,
\begin{equation}
\varphi(\mathbf{x})=(2\pi)^{-3/2}\int(2\omega_{\mathbf{k}})^{-1/2}[a(\mathbf{k})+a^{\dag}(-\mathbf{k})]
exp(i\mathbf{k}\mathbf{x})d\mathbf{k}, \label{3.49}
\end{equation}
\begin{equation}
\pi(\mathbf{x})=-i(2\pi)^{-3/2}\int(\omega_{\mathbf{k}}/2)^{1/2}[a(\mathbf{k})-a^{\dag}(-\mathbf{k})]
exp(i\mathbf{k}\mathbf{x})d\mathbf{k}, \label{3.50}
\end{equation}
\[
\psi(\mathbf{x})=(2\pi)^{-3/2}\int(m/E{\mathbf{p}})^{1/2}\sum_{\mu}[u(\mathbf{p}\mu)b(\mathbf{p}\mu)
\]
\begin{equation}
+v(-\mathbf{p}\mu)d^{\dag}(-\mathbf{p}\mu)]exp(i\mathbf{p}\mathbf{x})d\mathbf{p}.
\label{3.51}
\end{equation}

Substituting (\ref{3.45}) into (\ref{3.42}), we find
\[
\ \mathbf{N}=\mathbf{N}_{F}+\mathbf{N}_{I}
\]
with
\[
\
\mathbf{N}_{F}=\mathbf{N}_{ferm}+\mathbf{N}_{\pi}=-\int\mathbf{x}H_{ferm}(\mathbf{x})d\mathbf{x}
-\int\mathbf{x}H_{\pi}(\mathbf{x})d\mathbf{x}
\]
and
\[
\ \mathbf{N}_{I}=-\int\mathbf{x}H_{I}(\mathbf{x})d\mathbf{x}.
\]

Now, taking into account the transformation properties of the
fermion field $\psi(x)$ and the pion field $\varphi(x)$ with
respect to $\Pi$, it is readily seen that in the D picture density
(\ref{3.45}) is a scalar, i.e.,
\begin{equation}
U_{F}(\Lambda,b)H_{SCH}(x)U_{F}^{-1}(\Lambda,b)= H_{SCH}(\Lambda
x+b),
 \label{3.52}
\end{equation}
so
\begin{equation}
U_{F}(\Lambda,b)H_{I}(x)U_{F}^{-1}(\Lambda,b)= H_{I}(\Lambda x+b).
 \label{3.53}
\end{equation}
Just such a property has been used for that example on p.7.

It is well known (see, e.g., Sect. 5.1 in \cite{WeinbergBook1995})
that for a large class of theories the property (\ref{3.53}) with
the corresponding interaction densities $H_{I}(x)$, being
supplemented by the condition
\begin{equation}
[H_{I}(x'),H_{I}(x)]=0 \,\,\, for \,\,\, (x'-x)^{2}\leq0,
 \label{3.54}
\end{equation}
plays a crucial role in ensuring the RI of the $S$-matrix.
Appendix A is contained explicit expressions of all free
generators for the $\pi N$ system and tests for them to be
satisfied the Poincar\'{e} algebra.

\subsection{An algebraic approach within the Hamiltonian formalism}

As mentioned above, we are addressing those theories that start
from a total Hamiltonian (\ref{2.12}) with the interaction
$H_{I}=\int H_{I}(\mathbf{x})d\mathbf{x}$ whose density is sum
(\ref{1.7}) so
\begin{equation}
H_{I}=H_{sc}+H_{nsc}\equiv \int H_{sc}(\mathbf{x})d\mathbf{x}+\int
H_{nsc}(\mathbf{x})d\mathbf{x}.
 \label{3.55}
\end{equation}
It means that consideration in Subsec. 3.1, where only the density
in the first integral has the property (\ref{3.53}), i.e.,
\begin{equation}
U_{F}(\Lambda,b)H_{sc}(x)U_{F}^{-1}(\Lambda,b)= H_{sc}(\Lambda
x+b).
 \label{3.56}
\end{equation}

Then, taking into account that the first relation (\ref{2.11}) is
equivalent to the equality
\begin{equation}
[\mathbf{N}_{F},H_{I}]=[H,\mathbf{N}_{I}], \label{3.57}
\end{equation}
we will evaluate its l.h.s.. In this connection, let us regard the
operator
\begin{equation}
H_{sc}(t)=\int H_{sc}(x)d\mathbf{x} \label{3.58}
\end{equation}
and its similarity transformation
\begin{equation}
e^{i\mathbf{\beta}\mathbf{N}_{F}}H_{sc}(t)e^{-i\mathbf{\beta}\mathbf{N}_{F}}=\int
H_{sc}(L(\mathbf{\beta})x)d\mathbf{x}, \label{3.59}
\end{equation}
where $L(\mathbf{\beta})$ is any Lorentz boost with the parameters
$\mathbf{\beta}=(\beta^{1},\beta^{2},\beta^{3})$.

From (\ref{3.59}) it follows that
\begin{equation}
ie^{i\beta^{1}N_{F}^{1}}[N_{F}^{1},H_{sc}(t)]e^{-i\beta^{1}N_{F}^{1}}=
\frac{\partial}{\partial\beta^{1}}\int
H_{sc}(L(\beta^{1})x)d\mathbf{x}, \label{3.60}
\end{equation}
whence, for instance,
\[
\
i[N_{F}^{1},H_{sc}(t)]=\lim_{\beta^{1}\rightarrow0}\frac{\partial}{\partial\beta^{1}}\int
H_{sc}(t-\beta^{1}x^{1},x^{1}-\beta^{1}t,x^{2},x^{3})d\mathbf{x}
\]
\begin{equation}
=-\int(t\frac{\partial}{\partial
x^{1}}H_{sc}(x)+x^{1}\frac{\partial}{\partial
t}H_{sc}(x))d\mathbf{x}, \label{3.61}
\end{equation}
since for the infinitesimal boost
\[
\
L(\mathbf{\beta})x=(t-\mathbf{\beta}\mathbf{x},\mathbf{x}-\mathbf{\beta}t).
\]
In turn, from (\ref{3.61}) we get
\[
\ [N_{F}^{1},H_{sc}]=i\lim_{t\rightarrow0}\int
(-it[P^{1},H_{sc}(x)]+ix^{1}[H_{F},H_{sc}(x)])d\mathbf{x},
\]
so
\begin{equation}
[\mathbf{N}_{F},H_{sc}]=-\int
\mathbf{x}[H_{F},H_{sc}(\mathbf{x})]d\mathbf{x}.
 \label{3.62}
\end{equation}
By using Eq. (\ref{3.62}) equality (\ref{3.57}) can be written as
\begin{equation}
-\int
\mathbf{x}[H_{F},H_{sc}(\mathbf{x})]d\mathbf{x}=[H_{F},\mathbf{N}_{I}]+[H_{I},\mathbf{N}_{I}]+[H_{nsc},\mathbf{N}_{F}].
\label{3.63}
\end{equation}
Evidently, this equation is fulfilled if we put
\begin{equation}
\mathbf{N}_{I}=\mathbf{N}_{B}\equiv-\int
\mathbf{x}H_{sc}(\mathbf{x})d\mathbf{x}
 \label{3.64}
\end{equation}
and
\begin{equation}
[H_{sc},\mathbf{N}_{I}]=-\int \mathbf{x}d\mathbf{x}\int
d\mathbf{x}'[H_{sc}(\mathbf{x}'),H_{sc}(\mathbf{x})]=[\mathbf{N}_{F}+\mathbf{N}_{I},H_{nsc}]
 \label{3.65}
\end{equation}
or
\[
\int d\mathbf{x}\int
d\mathbf{x}'(\mathbf{x}'-\mathbf{x})[H_{sc}(\mathbf{x}'),H_{sc}(\mathbf{x})]
\]
\begin{equation}
=\int \mathbf{x}d\mathbf{x}\int
d\mathbf{x}'[H_{nsc}(\mathbf{x}'),H_{F}(\mathbf{x})+H_{sc}(\mathbf{x})].
\label{3.66}
\end{equation}
 In a model with $H_{nsc}=0$ the latter reduces to
\begin{equation}
\int
e^{-i\mathbf{P}\mathbf{X}}\mathbf{I}e^{i\mathbf{P}\mathbf{X}}d\mathbf{X}=0,
\label{3.67}
\end{equation}
where
\begin{equation}
\mathbf{I}=\frac{1}{2}\int
\mathbf{r}d\mathbf{r}[H_{sc}(\frac{1}{2}\mathbf{r}),H_{sc}(-\frac{1}{2}\mathbf{r})].
\label{3.68}
\end{equation}
By running again the way from Eq. (\ref{3.57}) to Eqs.
(\ref{3.67})-(\ref{3.68}) we see that the nonlinear commutation
(\ref{2.11})
\[
\ [H,\mathbf{N}]=i\mathbf{P}
\]
will take place once along with the Belinfante-type relation
(\ref{3.64}) the interaction density meets the condition
\begin{equation}
\int
\mathbf{r}d\mathbf{r}[H_{sc}(\frac{1}{2}\mathbf{r}),H_{sc}(-\frac{1}{2}\mathbf{r})]=0.
\label{3.69}
\end{equation}
One should note that we have arrived to Eq. (\ref{3.64}) being
inside the Poincar\`{e} algebra itself without addressing the
Noether integrals, these stepping stones of the Lagrangian
formalism. In the context, we would like to stress that the
condition (\ref{3.69}) is weaker compared to the constraint
\begin{equation}
[H_{sc}(\frac{1}{2}\mathbf{r}),H_{sc}(-\frac{1}{2}\mathbf{r})]=0
\label{3.70}
\end{equation}
imposed for all $\mathbf{r}$ excepting, may be, the point
$\mathbf{r}=0$. But we recall it as a special case of the
microcausality requirement that is realized in local field models.
Beyond such models, as it will be shown in Appendix B, Eqs.
(\ref{3.64}) and (\ref{3.57}) may be incompatible. It makes us
seek an alternative to assumption (\ref{3.64}) in our attempts to
meet Eq. (\ref{3.63}).

At this point, we put $\mathbf{N}_{I}=\mathbf{N}_{B}+\mathbf{D}$
to get in the framework of our consideration the relationship
\begin{equation}
[H_{F},\mathbf{D}]=[\mathbf{N}_{B}+\mathbf{D},H_{sc}]+
[\mathbf{N}_{F}+\mathbf{N}_{B}+\mathbf{D},H_{nsc}],
 \label{3.71}
\end{equation}
that replaces the commutator $[H,\mathbf{N}]=i\mathbf{P}$ and
determines the displacement $\mathbf{D}$.

Assuming that the scalar density $H_{sc}(\mathbf{x})$ is of the
first order in coupling constants involved and putting
\begin{equation}
H_{nsc}(\mathbf{x})=\sum_{p=2}^\infty H_{nsc}^{(p)}(\mathbf{x}),
 \label{3.72}
\end{equation}
we will search the operator $\mathbf{D}$ in the form
\begin{equation}
\mathbf{D}=\sum_{p=2}^\infty \mathbf{D}^{(p)},
 \label{3.73}
\end{equation}
i.e., as a perturbation expansion in powers of the interaction
$H_{sc}$. Here the label $(p)$ denotes the pth order in these
constants. By the way, one should keep in mind that the terms in
the r.h.s. of  Eq. (\ref{3.72}) are usually associated with
perturbation series for mass and vertex counterterms. Evidently,
their incorporation may affect the corresponding higher-order
contributions with $p\geq2$ to the boost. In this context, to
comprise different situations of practical interest let us
consider field models in which
\[
H_{nsc}(\mathbf{x})=V_{nsc}(\mathbf{x})+V_{ren}(\mathbf{x})
\]
with a nonscalar interaction
\[
V_{nsc}=\int V_{nsc}(\mathbf{x})d\mathbf{x}
\]
and some "renormalization" contribution
\[
V_{ren}=\int V_{ren}(\mathbf{x})d\mathbf{x}.
\]
The latter may be scalar or not. Of course, such a division of
$H_{nsc}(\mathbf{x})$ can be done at the beginning in Eq.
(\ref{3.55}). But the scheme, introduced here, seems to us more
flexible.

By substituting the expansions (\ref{3.72}) and (\ref{3.73}) into
Eq. (\ref{3.71}) we get the chain of relations
\begin{equation}
[H_{F},\mathbf{D}^{(2)}]= [\mathbf{N}_{F},H_{nsc}^{(2)}] +
[\mathbf{N}_{B},H_{sc}],
 \label{3.74}
\end{equation}
\begin{equation}
[H_{F},\mathbf{D}^{(3)}]= [\mathbf{N}_{F},H_{nsc}^{(3)}]
+[\mathbf{D}^{(2)},H_{sc}]+[\mathbf{N}_{B},H_{nsc}^{(2)}] ,
 \label{3.75}
\end{equation}
\begin{equation}
[H_{F},\mathbf{D}^{(p)}]=[\mathbf{N}_{F},H_{nsc}^{(p)}]+[\mathbf{N}_{B},H_{nsc}^{(p-1)}]
+[\mathbf{D}^{(p-1)},H_{sc}]+[\mathbf{D},H_{nsc}]^{(p)},
 \label{3.76}
\end{equation}
\[
(p = 4,5, \ldots )
\]
for a recursive finding of the operators $\mathbf{D}^{(p)}$
$(p=2,3,...)$.

Further, after such substitutions into the commutators
\[
[P_{k},N_{j}]=i\delta_{kj}H,
\]
\[
[J_{k},N_{j}]=i\varepsilon_{kjl}N_{l}
\]
and
\[
[N_{k},N_{j}]=-i\varepsilon_{kjl}J_{l} \footnote{The remaining
Poincar\'{e} commutations are fulfilled once one deals with any
rotationally and translationally invariant theory}
\]
we deduce, respectively, the following relations:
\begin{equation}
[P_{k},D^{(p)}_{j}]=i\delta_{kj}H_{nsc}^{(p)}\,\,\,\,\,\,(p=2,3,...)
 \label{3.77}
\end{equation}
from
\begin{equation}
[P_{k},D_{j}]=i\delta_{kj}H_{nsc},
 \label{3.78}
\end{equation}
\begin{equation}
[J_{k},D^{(p)}_{j}]=i\varepsilon_{kjl}D_{l}^{(p)}
 \label{3.79}
\end{equation}
from
\begin{equation}
[J_{k},D_{j}]=i\varepsilon_{kjl}D_{l}
 \label{3.80}
\end{equation}
and
\begin{equation}
[N_{Fk},N_{Bj}]+[N_{Bk},N_{Fj}]=0,
 \label{3.81}
\end{equation}
\begin{equation}
[N_{Fk},D_{j}^{(2)}] + [D_{k}^{(2)},N_{Fj}]+[N_{Bk},N_{Bj}]= 0,
 \label{3.82}
\end{equation}
\begin{equation}
[N_{Fk},D_{j}^{(3)}]
+[D_{k}^{(3)},N_{Fj}]+[N_{Bk},D_{j}^{(2)}]+[D_{k}^{(2)},N_{Bj}] =
0,
 \label{3.83}
\end{equation}
\begin{equation}
[N_{Fk},D_{j}^{(p)}]+[D_{k}^{(p)},N_{Fj}]+[N_{Bk},D_{j}]^{(p)}+[D_{k},N_{Bj}]^{(p)}
+[D_{k},D_{j}]^{(p)}=0,
 \label{3.84}
\end{equation}
\[
(p = 4,5, \ldots )
\]
from
\begin{equation}
[N_{Fk},N_{Bj}+D_{j}]+[N_{Bk}+D_{k},N_{Fj}]+[N_{Bk}+D_{k},N_{Bj}+D_{j}]=0.
 \label{3.85}
\end{equation}
Now, keeping in mind an elegant method by Chandler
\cite{Chandler}, we invoke on the property (see \cite{Friedrichs})
of a formal solution $Y$ to the equation
\begin{equation}
[H_{F},Y]=X
 \label{3.86}
\end{equation}
to be any linear functional $F(X)$ of a given operator $X\neq0$.
In other words, it means that
\begin{equation}
[H_{F},F(X)]=X
 \label{3.87}
\end{equation}
with
$F(\lambda_{1}X_{1}+\lambda_{2}X_{2})=\lambda_{1}F(X_{1})+\lambda_{2}F(X_{2})$,
where $\lambda_{1}$ and $\lambda_{2}$ are arbitrary c-numbers. In
addition, one can see that
\begin{equation}
[H_{F},F(X)]=F([H_{F},X]).
 \label{3.88}
\end{equation}
Moreover, it turns out that
\begin{equation}
[\mathbf{P},F(X)]=F([\mathbf{P},X]),
 \label{3.89}
\end{equation}
\begin{equation}
[\mathbf{J},F(X)]=F([\mathbf{J},X]),
 \label{3.90}
\end{equation}
\begin{equation}
[\mathbf{N}_{F},F(X)]=F([\mathbf{N}_{F},X])+iF(F([\mathbf{P},X])).
 \label{3.91}
\end{equation}
In order to prove the relations let us employ the Jacobi identity
\begin{equation}
[A,[B,C]]+[C,[A,B]]+[B,[C,A]]=0
 \label{3.92}
\end{equation}
and write
\[
[\mathcal{O},[H_{F},F(X)]]=-[F(X),[\mathcal{O},H_{F}]]+[H_{F},[\mathcal{O},F(X)]]
\]
with some operator $\mathcal{O}$. Then
\begin{equation}
[\mathcal{O},F(X)]=F([\mathcal{O},X])+F([F(X),[\mathcal{O},H_{F}]]).
 \label{3.93}
\end{equation}
The formulae (\ref{3.89})-(\ref{3.91}) follow from Eq.
(\ref{3.93}) if one takes into account the Poincar\'{e}
commutators for the free generators $G_{F}$. We derive them here
after \cite{Chandler} when moving from the nonlinear commutation
(\ref{3.71}) to ensuring the RI as a whole. After this let us
verify all commutations (\ref{3.77})- (\ref{3.85}) when one uses
the solution $Y = F(X)$ to Eq. (\ref{3.86}).

First, with the help of (\ref{3.89}) we find from Eqs.
(\ref{3.74})- (\ref{3.75}),
\[
[P_{k},D^{(2)}_{j}]
=F([P_{k},[N_{Fj},H_{nsc}^{(2)}]])=F([[P_{k},N_{Fj}],H_{nsc}^{(2)}])
\]
\begin{equation}
=i\delta_{kj}F([H_{F},H_{nsc}^{(2)}]) = i\delta_{kj}H_{nsc}^{(2)},
 \label{3.94}
\end{equation}
\[
[P_{k},D^{(3)}_{j}]
=F([P_{k},[N_{Fj},H_{nsc}^{(3)}]])=F([[P_{k},N_{Fj}],H_{nsc}^{(3)}])
\]
\begin{equation}
=i\delta_{kj}F([H_{F},H_{nsc}^{(3)}]) = i\delta_{kj}H_{nsc}^{(3)}.
 \label{3.95}
\end{equation}
We have used the formulae $[P_{k},N_{Bj}] = i\delta_{kj}H_{sc}$
and $[P_{k},H_{sc}] = [P_{k},H_{nsc}] = 0 $. Analogously, one can
verify Eqs. (\ref{3.79}) with $p = 2,3$. Second, Eq. (\ref{3.81})
is trivial.

Third,
\[
[N_{Fk},D_{j}^{(2)}] + [D_{k}^{(2)},N_{Fj}]+[N_{Bk},N_{Bj}]
\]
\[
=F([N_{Fk},[N_{Bj},H_{sc}]])-F([N_{Fj},[N_{Bk},H_{sc}]])+[N_{Bk},N_{Bj}]
\]
\[
=-F([N_{Bj},[N_{Bk},H_{F}]])+F([N_{Bk},[N_{Bj},H_{F}]])+[N_{Bk},N_{Bj}]
\]
\begin{equation}
=F([H_{F},[N_{Bj},N_{Bk}]])+[N_{Bk},N_{Bj}]=0 \label{3.96}
\end{equation}
and
\[
[N_{Fk},D_{j}^{(3)}]
+[D_{k}^{(3)},N_{Fj}]+[N_{Bk},D_{j}^{(2)}]+[D_{k}^{(2)},N_{Bj}]
\]
\[
=-F([H_{sc},[N_{Fk},D_{j}^{(2)}]])-F([N_{Bk},[D_{j}^{(2)},H_{F}]])
\]
\[
-[N_{Bk},D_{j}^{(2)}]+ F([N_{Fk},[N_{Bj},H_{nsc}^{(2)}]])
\]
\[
+F([H_{sc},[N_{Fj},D_{k}^{(2)}]])+F([N_{Bj},[D_{k}^{(2)},H_{F}]])
\]
\[
+[N_{Bj},D_{k}^{(2)}]- F([N_{Fj},[N_{Bk},H_{nsc}^{(2)}]])
\]
\[
+[N_{Bk},D_{j}^{(2)}]+[D_{k}^{(2)},N_{Bj}]
\]
\[
=F([H_{sc},[N_{Bk},N_{Bj}]])+F([N_{Bk},[N_{Bj},H_{sc}]])
\]
\[
+F([N_{Bj},[H_{sc},N_{Bk}]])+F([H_{nsc}^{(2)},[N_{Fk},N_{Bj}]])
\]
\begin{equation}
+F([N_{Fk},[N_{Bj},H_{nsc}^{(2)}]])+F([N_{Bj},[H_{nsc}^{(2)},N_{Fk}]])=0.
\label{3.97}
\end{equation}
At last, Eqs. (\ref{3.77}) and Eqs. (\ref{3.79}) with $p \geq 3$
and Eqs. (\ref{3.84}) with $p \geq 4$ can be proved inductively.
One should emphasize that for these derivations we have again
addressed the strategy chosen in \cite {Chandler}. Unfortunately,
that approach by Chandler is either well forgotten or little
known. Therefore, we are trying to present an entire picture.
However, to be more constructive one needs to have a definite
realization of the functional $F(X)$. In this connection, we will
use the representation
\begin{equation}
Y=-i\lim_{\eta\rightarrow0+}\int_{0}^{\infty}X(t)e^{-\eta t}dt
 \label{3.98}
\end{equation}
of the operator $Y$ that enters the equation (\ref{3.86}). The
existence proof for such a solution is sufficiently delicate (see
discussion in Appendix A of Ref. \cite{SheShi01}). Of course, it
depends on the operator $X$. We shall come back to the point in
Subsec. 3.3 for a situation, where $[H_F, X] = 0$.

Henceforth, the ensuring of RI via Eqs. (\ref{3.71})--(\ref{3.76})
calls the way I.

\subsection{Comparison with other approaches. Application to a nonlocal field model}
There are different perturbative schemes to meet the Poincar\'{e}
algebra (at least, in its instant form after Dirac). One of them,
elaborated in \cite{KruegGloe99}, is based upon a simultaneous
blockdiagonalization of the field Hamiltonian and the boost
operators by using a development of the Okubo idea
\cite{GloeMuel81} and constructing the corresponding unitary
transformation in a perturbative way (see also Sect. 6 in
\cite{SheShi01}).

An entirely algebraic approach \cite{Kita 68} (see also
\cite{Chandler} and a private communication to A.S.) is most close
to that exposed in Subsec. 3.2. In fact, its departure point is to
apply a perturbation expansion of the commutation relations
(\ref{2.9})-(\ref{2.11}) inserting into them the series
\[
\ H_{I}=\sum_{p=1}^\infty H_{I}^{(p)}
\]
and
\[
\ \mathbf{N}_{I}=\sum_{p=1}^\infty \mathbf{N}_{I}^{(p)},
\]
\begin{equation}
[H_{F},\mathbf{N}_{I}^{(1)}]=[\mathbf{N}_{F},H_{I}^{(1)}],
 \label{3.99}
\end{equation}
\begin{equation}
[H_{F},\mathbf{N}_{I}^{(2)}]=[\mathbf{N}_{F},H_{I}^{(2)}]+[\mathbf{N}_{I}^{(1)},H_{I}^{(1)}],
 \label{3.100}
\end{equation}
\begin{equation}
[H_{F},
\mathbf{N}_{I}^{(3)}]=[\mathbf{N}_{F},H_{I}^{(3)}]+[\mathbf{N}_{I}^{(1)},H_{I}^{(2)}]+[\mathbf{N}_{I}^{(2)},H_{I}^{(1)}],
\label{3.101}
\end{equation}
\begin{equation}
[P^{i}, N_{I}^{(p)j}]=\delta_{ij}H_{I}^{(p)}  \qquad (p = 1,2,
\ldots ) \label{3.102}
\end{equation}
\[
\ldots \qquad \ldots \qquad \ldots \qquad \ldots
\]
The recursive procedure based upon Eqs.
(\ref{3.99})--(\ref{3.101}) will be referred to as the way II.

Now, making a comparison between I and II we obtain with the aid
of formula (\ref{3.98}) the lowest-order terms:
\begin{equation}
\mathbf{N}_{I}^{(1)} = \mathbf{N}_{B} =
-\int\mathbf{x}H_{I}^{(1)}(\mathbf{x})d\mathbf{x}=
-\int\mathbf{x}H_{sc}(\mathbf{x})d\mathbf{x}
 \label{3.103}
\end{equation}
and
\begin{equation}
\mathbf{N}_{I}^{(2)} =\mathbf{D}^{(2)}=
\Phi([\mathbf{N}_{F},H_{nsc}^{(2)}])
-i\lim_{\eta\rightarrow0+}\int_{0}^{\infty}[\mathbf{N}_{B}(t),H_{sc}(t)]e^{-\eta
t}dt \label{3.104}
\end{equation}
from Eq. (\ref{3.64}) and Eq. (\ref{3.74}) vs
\begin{equation}
\mathbf{N}_{I}^{(1)}
=-i\lim_{\eta\rightarrow0+}\int_{0}^{\infty}[\mathbf{N}_{F}(t),H_{sc}(t)]e^{-\eta
t}dt \label{3.105}
\end{equation}
and
\begin{equation}
\mathbf{N}_{I}^{(2)} =\Phi([\mathbf{N}_{F},H_{nsc}^{(2)}])
-i\lim_{\eta\rightarrow0+}\int_{0}^{\infty}[\mathbf{N}_{I}^{(1)}(t),H_{sc}(t)]e^{-\eta
t}dt \label{3.106}
\end{equation}
from Eq. (\ref{3.99}) and Eq. (\ref{3.100}), respectively. The
first terms in the r.h.s. of Eq. (\ref{3.104}) and Eq.
(\ref{3.106}) have been expressed through a linear functional
$\Phi(X)$ since its argument $X=[\mathbf{N}_{F},H_{nsc}^{(2)}]$,
in general, can embody a part that commutes with $H_{F}$ (see that
note below the recipe (\ref{3.98})).

It is easily seen that these relations give rise to identical
results since the commutator in the r.h.s. Eq. (\ref{3.105}) can
be written as (see Eq. (\ref{3.62}))

\begin{equation}
[\mathbf{N}_{F}(t), H_{sc}(t)] =-\int
\mathbf{x}d\mathbf{x}[H_{F},H_{sc}(t)]=[H_{F},\mathbf{N}_{B}(t)]
\label{3.107}
\end{equation}
or
\[
[\mathbf{N}_{F}(t), H_{sc}(t)] =-i\frac{d}{dt}\mathbf{N}_{B}(t),
\]
so
\[
\mathbf{N}_{I}^{(1)}=-i\lim_{\eta\rightarrow
0+}\int_{0}^{\infty}[\mathbf{N}_{F}(t), H_{sc}(t)]e^{-\eta t}dt
\]
\[
 =\mathbf{N}_{B}(0)- \lim_{\eta\rightarrow 0+}\eta\int_{0}^{\infty}
\mathbf{N}_{B}(t) e^{-\eta t}dt=\mathbf{N}_{B}.
\]
By assumption,
\[
\lim_{\eta\rightarrow 0+}\eta\int_{0}^{\infty} \mathbf{N}_{B}(t)
e^{-\eta t}dt = 0,
\]
that should be verified every time for a given model interaction.

Besides, if the condition (\ref{3.69}) takes place, the approach
II enables us to arrive to the same result as our approach does,
i.e., the Belinfante ansatz by Eq. (\ref{3.64}).

As mentioned, the latter is inherent in some local field theories.
Therefore, we would like to employ the way I when handling
nonlocal field models. Let us consider a system of "scalar
nucleons" (more precisely, charged spinless bosons) and neutral
scalar bosons (see, e.g., Chapter 1 in \cite{Gas}) with the
following interaction density (cf. \cite{GloeMuel81},
\cite{Shir2002}):
\begin{equation}
H_{I}(\mathbf{x})=V_{loc}(\mathbf{x})+V_{ren}(\mathbf{x}),
 \label{3.108}
\end{equation}
\begin{equation}
V_{loc}(\mathbf{x})=g\varphi_{s}(\mathbf{x}):\psi_{b}^{\dag}(\mathbf{x})\psi_{b}(\mathbf{x}):
 \label{3.109}
\end{equation}
and
\begin{equation}
V_{ren}(\mathbf{x})=\delta\mu_{s}:\varphi_{s}^{2}(\mathbf{x}):+
\delta\mu_{b}:\psi_{b}^{\dag}(\mathbf{x})\psi_{b}(\mathbf{x}):
\label{3.110}
\end{equation}
with the mass shifts
$\delta\mu_{s}=\frac{1}{2}(\mu_{0s}^{2}-\mu_{s}^{2})$($\delta\mu_{b}=(\mu_{0b}^{2}-\mu_{b}^{2})$).
In order to regard a nonlocal extension of this local model let us
substitute the expansions
\[
\varphi_{s}(\mathbf{x})=[2(2\pi)^{3}]^{-1/2}\int\frac{d\mathbf{k}}{\omega_{\mathbf{k}}}[a(k)+a^{\dag}(k_{-})]
e^{i\mathbf{k}\mathbf{x}},
\]
\[
\psi_{b}(\mathbf{x})=[2(2\pi)^{3}]^{-1/2}\int\frac{d\mathbf{p}}{E_{\mathbf{p}}}
[b(p)+d^{\dag}(p_{-})]e^{i\mathbf{p}\mathbf{x}}
\]
into Eqs. (\ref{3.109}) and (\ref{3.110}) to get
\[
V_{loc}(\mathbf{x})=g[2(2\pi)^{3}]^{-3/2}\int\frac{d\mathbf{p'}}{E_{\mathbf{p'}}}
\int\frac{d\mathbf{p}}{E_{\mathbf{p}}}\int\frac{d\mathbf{k}}{\omega_{\mathbf{k}}}
e^{-i\mathbf{p'}\mathbf{x}+i\mathbf{p}\mathbf{x}+i\mathbf{k}\mathbf{x}}
\]
\begin{equation}
\times:[a(k)+a^{\dag}(k_{-})][b^{\dag}(p')+d(p'_{-})][b(p)+d^{\dag}(p_{-})]:
\label{3.111}
\end{equation}
and
\begin{equation}
V_{ren}(\mathbf{x})=\delta\mu_{s}(\mathbf{x})+\delta\mu_{b}(\mathbf{x})
\label{3.112}
\end{equation}
with
\begin{equation}
\delta\mu_{s}(\mathbf{x})=
\frac{\delta\mu_{s}}{2(2\pi)^3}\int\frac{d\mathbf{k'}}{\omega_{\mathbf{k'}}}
\int\frac{d\mathbf{k}}{\omega_{\mathbf{k}}}
:[a(k')+a^{\dag}(k'_{-})]e^{i\mathbf{k'}\mathbf{x}+i\mathbf{k}\mathbf{x}}[a(k)+a^{\dag}(k_{-})]:,
\label{3.113}
\end{equation}
\begin{equation}
\delta\mu_{b}(\mathbf{x})=
\frac{\delta\mu_{b}}{2(2\pi)^3}\int\frac{d\mathbf{p'}}{E_{\mathbf{p'}}}
\int\frac{d\mathbf{p}}{E_{\mathbf{p}}}
:[b^{\dag}(p')+d(p'_{-})]e^{-i\mathbf{p'}\mathbf{x}+i\mathbf{p}\mathbf{x}}[b(p)+d^{\dag}(p_{-})]:.
\label{3.114}
\end{equation}

The interaction operator itself
\[
H_{I}=\int H_{I}(\mathbf{x})d\mathbf{x}=V_{loc}+V_{ren},
\]
\[
V_{loc}=\int
V_{loc}(\mathbf{x})d\mathbf{x}=\frac{g}{2[2(2\pi)^3]^{1/2}}
\int\frac{d\mathbf{p'}}{E_{\mathbf{p'}}}
\int\frac{d\mathbf{p}}{E_{\mathbf{p}}}
\int\frac{d\mathbf{k}}{\omega_{\mathbf{k}}}
\delta(\mathbf{p'}-\mathbf{p}-\mathbf{k})
\]
\begin{equation}
\times
a(k):[b^{\dag}(p')b(p)+b^{\dag}(p')d^{\dag}(p_{-})+d(p'_{-})b(p)+d(p'_{-})d^{\dag}(p_{-})]:+H.c.,
\label{3.115}
\end{equation}
\begin{equation}
V_{ren}=\int[\delta\mu_{s}(\mathbf{x})+\delta\mu_{b}(\mathbf{x})]d\mathbf{x}.
\label{3.116}
\end{equation}

Let us consider its nonlocal extension
\begin{equation}
H_{I}=V_{nloc}+M_{s}+M_{b},
 \label{3.117}
\end{equation}
where in accordance with the representation (\ref{1.3}) we
introduce the following normally-ordered structures:
\[
V_{nloc}=\int V_{nloc}(\mathbf{x})d\mathbf{x}=
\int\frac{d\mathbf{p'}}{E_{\mathbf{p'}}}
\int\frac{d\mathbf{p}}{E_{\mathbf{p}}}
\int\frac{d\mathbf{k}}{\omega_{\mathbf{k}}}
\]
\[
\times\{\delta(\mathbf{p'}-\mathbf{p}-\mathbf{k})g_{11}(p',p,k)b^{\dag}(p')b(p)
+\delta(\mathbf{p'}+\mathbf{p}-\mathbf{k})g_{12}(p',p,k)b^{\dag}(p')d^{\dag}(p)
\]
\[
+\delta(\mathbf{p'}+\mathbf{p}+\mathbf{k})g_{21}(p',p,k)d(p')b(p)
\]
\begin{equation}
+\delta(\mathbf{p'}-\mathbf{p}-\mathbf{k})g_{22}(p',p,k)d^{\dag}(p')d(p)\}a(k)+H.c.
\label{3.118}
\end{equation}
or in more compact form
\[
V_{nloc}=V_{b}+V_{b}^{\dag},
\]
\begin{equation}
V_{b}=\int V_{b}(\mathbf{x})d\mathbf{x}
=\int\frac{d\mathbf{k}}{\omega_{\mathbf{k}}}:F^{\dag}_{b}G(k)F_{b}:a(k),
\label{3.119}
\end{equation}
where
\[
V_{b}(\mathbf{x})=\int\frac{d\mathbf{k}}{\omega_{\mathbf{k}}}e^{i\mathbf{k}\mathbf{x}}
:F^{\dag}_{b}G_{k}(\mathbf{x})F_{b}:a(k)
\]
with
\[
\{G_{k}(\mathbf{x})\}_{\varepsilon'\varepsilon}=
\frac{1}{(2\pi)^3}\bar{g}_{\varepsilon'\varepsilon}(p',p,k)
e^{i((-1)^{\varepsilon'}\mathbf{p'}-(-1)^{\varepsilon}\mathbf{p})\mathbf{x}},
\]
while the operators $M_{s}$ and $M_{b}$ will be given below.

Adopting the convention
\[
[b^{\dag}(p'), d(p')] \left[
\begin{array}{ll}
X_{11}(p',p) & X_{12}(p',p)   \\
X_{21}(p',p) & X_{22}(p',p)
\end{array}
\right]
 \left[
\begin{array}{l}
b(p) \\
d^{\dag}(p)
\end{array}
\right]
\]
\begin{equation}
=F^{\dag}_{\varepsilon'}(p')X_{\varepsilon'\varepsilon}(p',p)F_{\varepsilon}(p)
\equiv F^{\dag}_{b}(p')X(p',p)F_{b}(p) \label{3.120}
\end{equation}
for any $2\times2$ matrix $X(p',p)$ and the column
\[
\ F_{b}(p)= \left[
\begin{array}{l}
b(p) \\
d^{\dag}(p)
\end{array}
\right]
 \ \equiv \left[
\begin{array}{l}
F_{1}(p) \\
F_{2}(p)
\end{array}
\right]
\]
(cf. formula (A.8) in \cite{SheShi01}), sometimes it is convenient
to proceed with
\[
\int\frac{d\mathbf{p'}}{E_{\mathbf{p'}}}\int\frac{d\mathbf{p}}{E_{\mathbf{p}}}
F^{\dag}_{b}(p')X(p',p)F_{b}(p)\equiv F^{\dag}_{b}XF_{b}.
\]
In this context the matrix $G(k)$ in Eq. (\ref{3.119}) is composed
of the elements
\begin{equation}
G_{\varepsilon'\varepsilon}(p',p,k)=\bar{g}_{\varepsilon'\varepsilon}(p',p,k)
\delta(\mathbf{k}+(-1)^{\varepsilon'}\mathbf{p'}-(-1)^{\varepsilon}\mathbf{p}),
\label{3.121}
\end{equation}
\[
(\varepsilon',\varepsilon=1,2)
\]
where $\bar{g}_{\varepsilon'\varepsilon}(p',p,k)$ coincide with
$g_{\varepsilon'\varepsilon}(p',p,k)$ except
$\bar{g}_{22}(p',p,k)=g_{22}(p,p',k)$.

It is implied that the operators $a(a^{\dag})$, $b(b^{\dag})$ and
$d(d^{\dag})$ meet the commutation relations
\begin{equation}
[a(k),a^{\dag}(k')]=k_{0}\delta(\mathbf{k}-\mathbf{k'}),
\label{3.122}
\end{equation}
\begin{equation}
[b(p),b^{\dag}(p')]=[d(p),d^{\dag}(p')]=p_{0}\delta(\mathbf{p}-\mathbf{p'})
\label{3.123}
\end{equation}
with all the remaining ones being zero. Here
$k_{0}=\omega_{\mathbf{k}}=\sqrt{\mathbf{k}^2+\mu^{2}_{s}}$
$(p_{0}=E_{\mathbf{p}}=\sqrt{\mathbf{p}^2+\mu^{2}_{b}})$ is the
energy of the neutral (charged) particle with the mass
$\mu_{s}(\mu_{b})$. By the way, from (\ref{3.123}) it follows that
\begin{equation}
[F_{\varepsilon'}(p'),F_{\varepsilon}^{\dag}(p)]=p_{0}\delta(\mathbf{p'}-\mathbf{p})\sigma_{\varepsilon'\varepsilon},
\label{3.124}
\end{equation}
where
$\sigma_{\varepsilon'\varepsilon}=(-1)^{\varepsilon-1}\delta_{\varepsilon'\varepsilon}$.

Furthermore, the creation/destruction operators have the
transformation properties like (\ref{2.19}). For example,
\begin{equation}
U_{F}(\Lambda)a(k)U_{F}^{-1}(\Lambda)=a(\Lambda k). \label{3.125}
\end{equation}
Therefore in the D picture
\begin{equation}
U_{F}(\Lambda)V_{loc}(x)U_{F}^{-1}(\Lambda)=V_{loc}(\Lambda x),
\label{3.126}
\end{equation}
i.e., the interaction density $V_{loc}(x)$ is a Lorentz scalar.

For our nonlocal model we will retain the property assuming that
 \begin{equation}
U_{F}(\Lambda)V_{nloc}(x)U_{F}^{-1}(\Lambda)=V_{nloc}(\Lambda x).
\label{3.127}
\end{equation}
 It is readily seen that this relation holds if the coefficients $g_{\varepsilon'\varepsilon}$ meet the condition
 \begin{equation}
g_{\varepsilon'\varepsilon}(\Lambda p',\Lambda p,\Lambda
k)=g_{\varepsilon'\varepsilon}(p',p,k).
 \label{3.128}
\end{equation}

On the mass shells with $p'^{2}=p^{2}=\mu_{b}^{2}$ and
$k^{2}=\mu_{s}^{2}$ the latter means that the functions
$g_{\varepsilon'\varepsilon}(p',p,k)$ can depend only upon the
invariants $p'p$, $p'k$ and $pk$.

The transition from $V_{loc}$ to $V_{nloc}$ can be interpreted as
an endeavor to regularize the theory. In the context, the
introduction of some cutoff functions
$g_{\varepsilon'\varepsilon}$ in momentum space is aimed at
removing ultraviolet divergences typical of local field models
with interactions like expression (\ref{3.109}).

One should keep in mind that along with the requirement
(\ref{3.128}) these cutoffs are subject to other constraints
imposed by different symmetries. For example, the tacit invariance
of the hermitian operator (\ref{3.118}) with respect to: i) space
inversion $\mathcal{P}$; ii) time reversal $\mathcal{T}$ and iii)
charge conjugation $\mathcal{C}$ yields the relations
\begin{equation}
g_{\varepsilon'\varepsilon}(p',p,k)=g_{\varepsilon'\varepsilon}(p,p',k),
 \,\,\,\,\varepsilon' \neq \varepsilon
\label{3.129}
\end{equation}
\begin{equation}
g_{\varepsilon'\varepsilon}(p',p,k)=g_{\varepsilon'\varepsilon}(p'_{-},p_{-},k_{-}),
\label{3.130}
\end{equation}
\begin{equation}
g_{11}(p',p,k)=g_{22}(p',p,k),
 \label{3.131}
\end{equation}
which can be derived assuming (see, e.g., Subsec. 5.2 in
\cite{WeinbergBook1995}) the following properties
\begin{equation}
\mathcal{P}a(\mathbf{k})\mathcal{P}^{-1}=a(-\mathbf{k}),\,\,\,
 \mathcal{P}b(\mathbf{p})\mathcal{P}^{-1}=b(-\mathbf{p}),\,\,\,
\mathcal{P}d(\mathbf{p})\mathcal{P}^{-1}= d(-\mathbf{p}),
\label{3.132}
\end{equation}
\begin{equation}
\mathcal{T}a(\mathbf{k})\mathcal{T}^{-1}=a(-\mathbf{k}),\,\,\,
 \mathcal{T}b(\mathbf{p})\mathcal{T}^{-1}=b(-\mathbf{p}),\,\,\,
\mathcal{T}d(\mathbf{p})\mathcal{T}^{-1}=d(-\mathbf{p}),
\label{3.133}
\end{equation}
\begin{equation}
\mathcal{C}a(\mathbf{k})\mathcal{C}^{-1}=a(\mathbf{k}),\,\,\,
\mathcal{C}b(\mathbf{p})\mathcal{C}^{-1}=d(\mathbf{p}),\,\,\,
\mathcal{C}d(\mathbf{p})\mathcal{C}^{-1}=b(\mathbf{p}),
\label{3.134}
\end{equation}
$\forall$ $\mathbf{p}$ and $\mathbf{k}$.

As to constructing the "mass renormalization terms" \footnote{We
will confine ourselves to the consideration of such terms. Of
course, the so-called charge and wave function counterterms can be
included too to be cancelled then by the $g^{3}$-order
contributions (cf. \cite{KoFro}) starting from the commutator
$\frac{1}{3}[R[R,V_{bad}]]$ in expansion (\ref{4.159})} $M_{s}$
and $M_{b}$ we note that within the clothing procedure exposed in
the next section they can be represented in the form:
\begin{equation}
M_{s}=\int\frac{d\mathbf{k}}{\omega_{\mathbf{k}}^2}\{m_{1}(k)a^{\dag}(k)a(k)+m_{2}(k)[a^{\dag}(k)a^{\dag}(k_{-})
+a(k)a(k_{-})]\} \label{3.135}
\end{equation}
and
\[
M_{b}=\int\frac{d\mathbf{p}}{E_{\mathbf{p}}^2}\{m_{11}(p)b^{\dag}(p)b(p)+
m_{12}(p)b^{\dag}(p)d^{\dag}(p_{-})
\]
\begin{equation}
+m_{21}(p)b(p)d(p_{-})+m_{22}(p)d^{\dag}(p)d(p)\}, \label{3.136}
\end{equation}
where the coefficients $m_{1,2}(k)$ and
$m_{\varepsilon'\varepsilon}(p)$, being for the time unknown, may
be momentum dependent. Of course, the latter (for simplicity,
real) should be symmetrical, i.e. $m_{12}(p)=m_{21}(p)$, to ensure
the hermiticity of $M_{b}$.

 Now, in order to derive the
corresponding lowest-order contributions to the boost operator for
our nonlocal model, we find using Eqs.
(\ref{3.103})-(\ref{3.104}),
\begin{equation}
\mathbf{N}_{I}^{(1)} = \mathbf{N}_{B} =-\int
\mathbf{x}V_{nloc}(\mathbf{x})d\mathbf{x}, \label{3.137}
\end{equation}
\begin{equation}
\mathbf{N}_{I}^{(2)}=\mathbf{D}^{(2)} =
\Phi([\mathbf{N}_{F},H_{nsc}^{(2)}])
-i\lim_{\eta\rightarrow0+}\int_{0}^{\infty}[\mathbf{N}_{B}(t),V_{nloc}(t)]e^{-\eta
t}dt, \label{3.138}
\end{equation}
\[
\mathbf{N}_{I}^{(3)}=\mathbf{D}^{(3)}
=-i\lim_{\eta\rightarrow0+}\int_{0}^{\infty}[\mathbf{N}_{B}(t),H_{nsc}^{(2)}(t)]e^{-\eta
t}dt
\]
\begin{equation}
-i\lim_{\eta\rightarrow0+}\int_{0}^{\infty}[\mathbf{D}^{(2)}(t),V_{nloc}(t)]e^{-\eta
t}dt
-i\lim_{\eta\rightarrow0+}\int_{0}^{\infty}[\mathbf{N}_{F}(t),H_{nsc}^{(3)}(t)]e^{-\eta
t}dt, \label{3.139}
\end{equation}
\[
\ldots \qquad \ldots \qquad \ldots \qquad \ldots
\]
where
$H_{nsc}^{(2)}(t)=exp(iH_{F}t)(M_{s}^{(2)}+M_{b}^{(2)})exp(-iH_{F}t)$
with the leading-order contributions $M_{s}^{(2)}$ and
$M_{b}^{(2)}$ to the operators $M_{s}$ and $M_{b}$ that will be
given explicitly below. We will confine ourselves to the
evaluation of contributions $\mathbf{N}_{I}^{(1)}$ and
$\mathbf{N}_{I}^{(2)}$. It suffices to conceive of some
manifestations of the model nonlocality.

Thus, by handling relation (\ref{3.138}), we encounter commutator
\[
[\mathbf{N}_{B}(t),
V_{nloc}(t)]=-\int\mathbf{x}'[V_{nloc}(t,\mathbf{x}'),V_{nloc}(t,\mathbf{x})]d\mathbf{x}d\mathbf{x}'
\]
\begin{equation}
=-\frac{1}{2}\int d\mathbf{x}'\int
d\mathbf{x}(\mathbf{x}'-\mathbf{x})[V_{nloc}(t,\mathbf{x}'),V_{nloc}(t,\mathbf{x})]
 \label{3.140}
\end{equation}
with
\[
[V_{nloc}(t,\mathbf{x}'),V_{nloc}(t,\mathbf{x})]=
exp(iH_{F}t)[V_{nloc}(\mathbf{x}'),V_{nloc}(\mathbf{x})]exp(-iH_{F}t),
\]
where
\begin{equation}
[V_{nloc}(\mathbf{x}'),V_{nloc}(\mathbf{x})]=
 [V_{b}(\mathbf{x}'),V_{nloc}(\mathbf{x})]-H.c.
 \label{3.141}
\end{equation}
and
\begin{equation}
[V_{b}(\mathbf{x}'),V_{nloc}(\mathbf{x})]=
 [V_{b}(\mathbf{x}'),V_{b}(\mathbf{x})]+[V_{b}(\mathbf{x}'),V_{b}^{\dag}(\mathbf{x})].
 \label{3.142}
\end{equation}
The first term in the r.h.s. of (\ref{3.142}) is equal to
\[
[V_{b}(\mathbf{x}'),V_{b}(\mathbf{x})] =[(2\pi)^{6}]^{-1}
\int\frac{d\mathbf{k}}{\omega_{\mathbf{k}}}
\int\frac{d\mathbf{k}_{1}}{\omega_{\mathbf{k}_{1}}}
\int\frac{d\mathbf{p}'}{E_{\mathbf{p}'}}
\int\frac{d\mathbf{p}}{E_{\mathbf{p}}}
\int\frac{d\mathbf{q}}{E_{\mathbf{q}}}
\]
\[
\times
[e^{i((-1)^{\varepsilon'}\mathbf{p}'-(-1)^{\rho'}\mathbf{q}+\mathbf{k})\mathbf{x}'}
e^{i((-1)^{\rho}\mathbf{q}-(-1)^{\varepsilon}\mathbf{p}+\mathbf{k}_{1})\mathbf{x}}
\]
\[
-e^{i((-1)^{\varepsilon'}\mathbf{p}'-(-1)^{\rho'}\mathbf{q}+\mathbf{k})\mathbf{x}}
e^{i((-1)^{\rho}\mathbf{q}-(-1)^{\varepsilon}\mathbf{p}+\mathbf{k}_{1})\mathbf{x}'}]
\]
\begin{equation}
\times
F^{\dag}_{\varepsilon'}(p')\bar{g}_{\varepsilon'\rho'}(p',q,k)\sigma_{\rho'\rho}
\bar{g}_{\rho\varepsilon}(q,p,k_{1})F_{\varepsilon}(p)a(k)a(k_{1}).
\label{3.143}
\end{equation}
This matrix form can be derived using commutations
\begin{equation}
 [F_{\varepsilon'}(p'),F^{\dag}_{\varepsilon}(p)]=p_{0}\delta(\mathbf{p}'-\mathbf{p})\sigma_{\varepsilon'\varepsilon},
 \label{3.144}
\end{equation}
where
$\sigma_{\varepsilon'\varepsilon}=(-1)^{\varepsilon-1}\delta_{\varepsilon'\varepsilon}$.

In turn, we have
\[
F^{\dag}_{\varepsilon'}(p')\bar{g}_{\varepsilon'\rho'}(p',q,k)\sigma_{\rho'\rho}
\bar{g}_{\rho\varepsilon}(q,p,k_{1})F_{\varepsilon}(p)
\]
\[
=F^{\dag}_{1}(p')[\bar{g}_{11}(p',q,k)\bar{g}_{11}(q,p,k_{1})
-\bar{g}_{12}(p',q,k)\bar{g}_{21}(q,p,k_{1})]F_{1}(p)
\]
\[
+F^{\dag}_{1}(p')[\bar{g}_{11}(p',q,k)\bar{g}_{12}(q,p,k_{1})
-\bar{g}_{12}(p',q,k)\bar{g}_{22}(q,p,k_{1})]F_{2}(p)
\]
\[
+F^{\dag}_{2}(p')[\bar{g}_{21}(p',q,k)\bar{g}_{11}(q,p,k_{1})
-\bar{g}_{22}(p',q,k)\bar{g}_{21}(q,p,k_{1})]F_{1}(p)
\]
\[
+F^{\dag}_{2}(p')[\bar{g}_{21}(p',q,k)\bar{g}_{12}(q,p,k_{1})
-\bar{g}_{22}(p',q,k)\bar{g}_{22}(q,p,k_{1})]F_{2}(p).
\]
When
$\bar{g}_{11}(p',p,k)=\bar{g}_{12}(p',p,k)=\bar{g}_{21}(p',p,k)\equiv\bar{g}(p',p,k)$,
we get
\begin{equation}
[V_{b}(\mathbf{x}'),V_{b}(\mathbf{x})]=0 \label{3.145}
\end{equation}
and
\begin{equation}
[V_{b}(\mathbf{x}'),V_{b}^{\dag}(\mathbf{x})]-H.c.=0 \label{3.146}
\end{equation}
so
\begin{equation}
[V_{nloc}(\mathbf{x}'),V_{nloc}(\mathbf{x})]=0. \label{3.147}
\end{equation}
Then we obtain from Eq. (\ref{3.138})
\begin{equation}
\mathbf{D}^{(2)} = \Phi([\mathbf{N}_{F},M_{s}^{(2)}+M_{b}^{(2)}])
\label{3.148}
\end{equation}
and we see that even with relation (\ref{3.147}) reminiscent of
the well-known microcausality condition (cf. Eq. (\ref{3.54})) one
has to evaluate the displacement operator $\mathbf{D}$, if the
mass renormalization terms are inequal to zero. But the latter is
the case. Otherwise, we would come to some contradiction with Eq.
(\ref{3.74}) and Eq. (\ref{3.77}) (details see in Subsec. 4.2).

By using the formulae  (\ref{A.14}) and (\ref{A.15}) and taking
into account that to an accuracy of adding an arbitrary function
of $H_{F}$ the solution $Y$ to $[H_{F},Y]=X$ repeats the operator
structure of $X$, we arrive to the division
\begin{equation}
\mathbf{D}^{(2)}=\mathbf{D}_{con}^{(2)}+\mathbf{D}_{ncon}^{(2)},
 \label{3.149}
\end{equation}
where the particle-number-conserving and -nonconserving
contributions $\mathbf{D}^{(2)}_{con}$ and
$\mathbf{D}^{(2)}_{ncon}$ are determined by
\[
\mathbf{D}_{con}^{(2)}=\frac{i}{2}\int\frac{d\mathbf{k}'}{\omega_{\mathbf{k}'}}
\int\frac{d\mathbf{k}}{\omega_{\mathbf{k}}}(\omega_{\mathbf{k}'}\omega_{\mathbf{k}}+\mathbf{k}'\mathbf{k}+\mu_{s}^{2})
(\frac{m^{(2)}_{1}(k)}{\omega_{\mathbf{k}}}-\frac{m^{(2)}_{1}(k')}{\omega_{\mathbf{k}'}})
\]
\[
\times\frac{a^{\dag}(k')a(k)}{\omega_{\mathbf{k}'}-\omega_{\mathbf{k}}}
\frac{\partial}{\partial\mathbf{k}}\delta(\mathbf{k}-\mathbf{k}')
\]
\[
+\frac{i}{2}\int\frac{d\mathbf{p}'}{E_{\mathbf{p}'}}
\int\frac{d\mathbf{p}}{E_{\mathbf{p}}}(E_{\mathbf{p}'}E_{\mathbf{p}}+\mathbf{p}'\mathbf{p}+\mu_{b}^{2})
(\frac{m^{(2)}_{11}(p)}{E_{\mathbf{p}}}-\frac{m^{(2)}_{11}(p')}{E_{\mathbf{p}'}})
\]
\[
\times\frac{b^{\dag}(p')b(p)} {E_{\mathbf{p}'}-E_{\mathbf{p}}}
\frac{\partial}{\partial\mathbf{p}}\delta(\mathbf{p}-\mathbf{p}')
\]
\[
+\frac{i}{2}\int\frac{d\mathbf{p}'}{E_{\mathbf{p}'}}
\int\frac{d\mathbf{p}}{E_{\mathbf{p}}}(E_{\mathbf{p}'}E_{\mathbf{p}}+\mathbf{p}'\mathbf{p}+\mu_{b}^{2})
(\frac{m^{(2)}_{22}(p)}{E_{\mathbf{p}}}-\frac{m^{(2)}_{22}(p')}{E_{\mathbf{p}'}})
\]
\begin{equation}
\times\frac{d^{\dag}(p')d(p)} {E_{\mathbf{p}'}-E_{\mathbf{p}}}
\frac{\partial}{\partial\mathbf{p}}\delta(\mathbf{p}-\mathbf{p}')
 \label{3.150}
\end{equation}
and
\[
\mathbf{D}_{ncon}^{(2)}=i\int\frac{d\mathbf{k}'}{\omega_{\mathbf{k}'}}
\int\frac{d\mathbf{k}}{\omega_{\mathbf{k}}}m^{(2)}_{2}(k)
\frac{\omega_{\mathbf{k}'}\omega_{\mathbf{k}}+\mathbf{k}'\mathbf{k}+\mu_{s}^{2}}{\omega_{\mathbf{k}}}
\]
\[
\times\frac{a^{\dag}(k')a^{\dag}(k_{-})-a(k')a(k_{-})}{\omega_{\mathbf{k'}}+\omega_{\mathbf{k}}}
\frac{\partial}{\partial\mathbf{k}}\delta(\mathbf{k}-\mathbf{k}')
\]
\[
+i\int\frac{d\mathbf{p}'}{E_{\mathbf{p}'}}
\int\frac{d\mathbf{p}}{E_{\mathbf{p}}}m^{(2)}_{12}(p)
\frac{E_{\mathbf{p}'}E_{\mathbf{p}}+\mathbf{p}'\mathbf{p}+\mu_{b}^{2}}{E_{\mathbf{p}}}
\]
\begin{equation}
\times\frac{b^{\dag}(p')d^{\dag}(p_{-})-b(p')d(p_{-})}{E_{\mathbf{p'}}+E_{\mathbf{p}}}
\frac{\partial}{\partial\mathbf{p}}\delta(\mathbf{p}-\mathbf{p}').
\label{3.151}
\end{equation}
One should point out that the operator $\mathbf{D}_{con}^{(2)}$
stems from the structure
\[
[\mathbf{N}_{F},M_{s}^{(2)}+M_{b}^{(2)}]\sim
a^{\dag}a+b^{\dag}b+d^{\dag}d,
\]
which commutes with $H_{F}$.

\section{Boost operators for clothed particles}
As shown in \cite{SheShi01}, the Belinfante ansatz turns out to be
useful when constructing the Lorentz boosts in the CPR. Their
generator $\mathbf{N}\equiv\mathbf{N}(\alpha)$, being a function
of the primary operators $\{\alpha\}$ (such as $a^{\dag}(a)$,
$b^{\dag}(b)$ and $d^{\dag}(d)$ for the examples regarded above)
in the BPR, is expressed through the corresponding operators
$\{\alpha_{c}\}$ for particle creation and annihilation in the
CPR. The transition $\{\alpha\}\Longrightarrow\{\alpha_{c}\}$ is
implemented via the special unitary transformations
$W(\alpha)=W(\alpha_{c})$, viz.,
\begin{equation}
\alpha=W(\alpha_{c})\alpha_{c}W^{\dag}(\alpha_{c}),
 \label{4.152}
\end{equation}
satisfying certain physical requirements (details can be also
found in Refs. \cite{KorShe04}, \cite{KorCanShe}).

\subsection{Elimination of bad terms in generators of the Poincar\'{e} group}
A key point of the clothing procedure exposed in \cite{SheShi01}
is to remove the so-called bad terms\footnote{For example, such
terms $b_{c}^{\dag}b_{c}a_{c}^{\dag}$,
$b_{c}^{\dag}d_{c}^{\dag}a_{c}$,
$b_{c}^{\dag}d_{c}^{\dag}a_{c}^{\dag}$,
$d_{c}d_{c}^{\dag}a_{c}^{\dag}$ enter $V(\alpha_{c})$ determined
by Eq. (\ref{3.115}) after the replacement of the bare operators
in it by the clothed ones. These terms are removed together with
their Hermitian conjugate counterterms to retain the hermiticity
of  the similarity transformation (\ref{4.153})} from the
Hamiltonian
\begin{equation}
H\equiv
H(\alpha)=H_{F}(\alpha)+H_{I}(\alpha)=W(\alpha_{c})H(\alpha_{c})W^{\dag}(\alpha_{c})\equiv
K(\alpha_{c}),
 \label{4.153}
\end{equation}
more exactly, from a primary interaction $V(\alpha)$ that enters
$H_{I}(\alpha)=V(\alpha)+V_{ren}(\alpha)$ (cf., e.g., our nonlocal
model with $V_{nloc}=V(\alpha)$ and
$V_{ren}=V_{ren}(\alpha)=M_{s}(\alpha)+M_{b}(\alpha)$). By
definition, such terms prevent the physical vacuum
$|\Omega\rangle$ (the $H$ lowest eigenstate) and the
one-clothed-particle states
$|n\rangle_{c}=a^{\dag}_{c}(n)|\Omega\rangle$ to be the $H$
eigenvectors for all $n$ included. Here creation operators
$a^{\dag}_{c}(n)$ are clothed counterparts of those operators
$a^{\dag}(n)$ that are contained in expansion (\ref{1.2}). The bad
terms \footnote{A recursive scheme for successive eliminations of
such terms has been regarded in \cite{KorCanShe}} occur every time
when any normally ordered product
\[
a^{\dag}(1')a^{\dag}(2')...a^{\dag}(n_{C}')a(n_{A})...a(2)a(1)
\]
of the class [C.A] embodies, at least, one substructure which
belongs to one of the classes $[k.0]$ $(k=1,2,...)$ and $[k.1]$
$(k=0,1,...)$.

Therefore, in correspondence with the decomposition (\ref{3.55})
we have
\begin{equation}
H_{I}(\alpha)=\int
H_{I}(\mathbf{x})d\mathbf{x}=H_{sc}(\alpha)+H_{nsc}(\alpha),
 \label{4.154}
\end{equation}
\[
H_{sc(nsc)}(\alpha)=\int H_{sc(nsc)}(\mathbf{x})d\mathbf{x},
\]
assuming that
\[
H_{sc}(\alpha)=V_{bad}(\alpha)+V_{good}(\alpha)
\]
 to remove the bad part $V_{bad}$\footnote{Remind
that term "good", as an antithesis of "bad", is applied here to
those operators (e.g., of the class [k.2] with $k\geq2$) which
destroy both the no-clothed-particle state $\Omega$ and the
one-clothed-particle states} from the similarity transformation
\[
K(\alpha_{c})=W(\alpha_{c})[H_{F}(\alpha_{c})+H_{I}(\alpha_{c})]W^{\dag}(\alpha_{c})
\]
\begin{equation}
=W(\alpha_{c})[H_{F}(\alpha_{c})+V_{bad}(\alpha_{c})+V_{good}(\alpha_{c})+H_{nsc}(\alpha_{c})]W^{\dag}(\alpha_{c}).
 \label{4.155}
\end{equation}
For the unitary clothing transformation (UCT) $W=expR$ with
$R=-R^{\dag}$\footnote{Sometimes, for brevity, we omit evident
arguments} it is implied that we will eliminate the bad terms
$V_{bad}$ in the r.h.s. of
\[
K(\alpha_{c})=H_{F}(\alpha_{c})+V_{bad}(\alpha_{c})+
[R,H_{F}]+[R,V_{bad}]+\frac{1}{2}[R,[R,H_{F}]]
\]
\begin{equation}
+\frac{1}{2}[R,[R,V_{bad}]]+...+e^{R}V_{good}e^{-R}+e^{R}H_{nsc}e^{-R}
 \label{4.156}
\end{equation}
(cf. Eq. (2.19) in \cite{SheShi01}) by requiring that
\begin{equation}
[H_{F},R]=V_{bad}
 \label{4.157}
\end{equation}
for the operator $R$ of interest.

One should note that unlike the original clothing procedure
exposed in \cite{SheShi01}, \cite{KorCanShe} we eliminate here the
bad terms only from $H_{sc}$ interaction in spite of such terms
can appear in the nonscalar interaction as well. This preference
is relied upon the previous experience \cite{Bonn09} and
\cite{FBS2010} when applying the method of UCTs in the theory of
nucleon-nucleon scattering. Now we get the division
\begin{equation}
H=K(\alpha_{c})=K_{F}+K_{I}
 \label{4.158}
\end{equation}
with a new free part $K_{F}=H_{F}(\alpha_{c})\sim
a^{\dag}_{c}a_{c}$ and interaction
\[
K_{I}=V_{good}(\alpha_{c})+H_{nsc}(\alpha_{c})+[R,V_{good}]
\]
\begin{equation}
+\frac{1}{2}[R,V_{bad}]+[R,H_{nsc}]+\frac{1}{3}[R,[R,V_{bad}]]+...,
 \label{4.159}
\end{equation}
where the r.h.s. involves along with good terms other bad terms to
be removed via subsequent UCTs described in Subsec. 2.4 of
\cite{SheShi01} and Sec. 3 of \cite{KorCanShe}.

In parallel, we have
\begin{equation}
\mathbf{N}\equiv\mathbf{N}(\alpha)=\mathbf{N}_{F}(\alpha)+\mathbf{N}_{I}(\alpha)
=W(\alpha_{c})\mathbf{N}(\alpha_{c})W^{\dag}(\alpha_{c})\equiv\mathbf{B}(\alpha_{c})
 \label{4.160}
\end{equation}
or
\begin{equation}
\mathbf{B}(\alpha_{c})=
\mathbf{N}_{F}(\alpha_{c})+\mathbf{N}_{I}(\alpha_{c})+[R,\mathbf{N}_{F}]+[R,\mathbf{N}_{I}]+...,
 \label{4.161}
\end{equation}
where accordingly the division
\begin{equation}
\mathbf{N}_{I}=\mathbf{N}_{B}+\mathbf{D},
 \label{4.162}
\end{equation}
\[
\mathbf{N}_{B}=-\int
\mathbf{x}H_{sc}(\mathbf{x})d\mathbf{x}=\mathbf{N}_{bad}+\mathbf{N}_{good},
\]
Eq. (\ref{4.161}) can be rewritten as
\[
\mathbf{B}(\alpha_{c})=\mathbf{N}_{F}(\alpha_{c})+\mathbf{N}_{bad}(\alpha_{c})+
[R,\mathbf{N}_{F}]+[R,\mathbf{N}_{bad}]+
\frac{1}{2}[R,[R,\mathbf{N}_{F}]]
\]
\begin{equation}
+\frac{1}{2}[R,[R,\mathbf{N}_{bad}]]+...+e^{R}\mathbf{N}_{good}e^{-R}+e^{R}\mathbf{D}e^{-R}.
 \label{4.163}
\end{equation}
But it turns out (see the proof of Eq. (3.26) in \cite{SheShi01})
that if $R$ meets the condition (\ref{4.157}), then
\begin{equation}
[\mathbf{N}_{F},R]=\mathbf{N}_{bad}=-\int\mathbf{x}V_{bad}(\mathbf{x})d\mathbf{x}
 \label{4.164}
\end{equation}
so the boost generators in the CPR can be written likely Eq.
(\ref{4.158}),
\begin{equation}
\mathbf{N}=\mathbf{B}(\alpha_{c})=\mathbf{B}_{F}+\mathbf{B}_{I},
 \label{4.165}
\end{equation}
where $\mathbf{B}_{F}=\mathbf{N}_{F}(\alpha_{c})$ is the boost
operator for noninteracting clothed particles while
$\mathbf{B}_{I}$ includes the contributions induced by
interactions between them
\[
\mathbf{B}_{I}=\mathbf{N}_{good}(\alpha_{c})+\mathbf{D}(\alpha_{c})+[R,\mathbf{N}_{good}]
\]
\begin{equation}
+\frac{1}{2}[R,\mathbf{N}_{bad}]+[R,\mathbf{D}]+\frac{1}{3}[R,[R,\mathbf{N}_{bad}]]+...
 \label{4.166}
\end{equation}
One should note that in formulae (\ref{4.159}) and (\ref{4.166})
we are focused upon the $R$-commutations with the first-eliminated
interaction $V_{bad}$. As shown in \cite{SheShi01}, the brackets,
on the one hand, yield new interactions responsible for different
physical processes and, on the other hand, cancel (as a recipe)
the mass and other counterterms that stem from
$H_{nsc}(\alpha_{c})$ and $\mathbf{D}(\alpha_{c})$. Such a
cancellation will be regarded in the next subsection.

But at this point we will come back to our model with
$V_{bad}=V_{nloc}$, $V_{good}=0$ and $R=R_{nloc}$ to calculate the
simplest commutator $[R_{nloc},V_{nloc}]$ in which accordingly
condition (\ref{4.157}) the clothing operator $R_{nloc}$ is
determined by
\begin{equation}
[H_{F},R_{nloc}]=V_{nloc}.
 \label{4.167}
\end{equation}
From the equation it follows (cf. Appendix A in \cite{SheShi01})
that its solution can be given by
\begin{equation}
R_{nloc}=\int\frac{d\mathbf{k}}{\omega_{\mathbf{k}}}:F^{\dag}_{b}R(k)F_{b}:a(k)-H.c.
=\mathcal{R}_{nloc}-\mathcal{R}_{nloc}^{\dag}.
 \label{4.168}
\end{equation}
The matrix $R(k)$ is composed of the elements
\begin{equation}
R_{\varepsilon'\varepsilon}(p',p,k)=-
\frac{\bar{g}_{\varepsilon'\varepsilon}(p',p,k)}
{\omega_{\mathbf{k}}+(-1)^{\varepsilon'}E_{\mathbf{p'}}-(-1)^{\varepsilon}E_{\mathbf{p}}}
\delta(\mathbf{k}+(-1)^{\varepsilon'}\mathbf{p'}-(-1)^{\varepsilon}\mathbf{p}).
\label{4.169}
\end{equation}
\[
(\varepsilon',\varepsilon=1,2)
\]
Such a solution is valid if $\mu_{s}<2\mu_{b}$. In other words,
under such an inequality the operator $R_{nloc}$ has the same
structure as $V_{nloc}$ itself. Then, all we need is to evaluate
\begin{equation}
[R_{nloc},V_{nloc}]=[\mathcal{R}_{nloc}-\mathcal{R}_{nloc}^{\dag},V_{nloc}]=
[\mathcal{R}_{nloc},V_{nloc}]+H.c., \label{4.170}
\end{equation}
where accordingly (\ref{3.119})
\begin{equation}
[\mathcal{R}_{nloc},V_{nloc}]=[\mathcal{R}_{nloc},V_{b}]+[\mathcal{R}_{nloc},V_{b}^{\dag}].
\label{4.171}
\end{equation}
Further, using Eqs. (\ref{3.119}), (\ref{4.168}) and identity
(\ref{B.10}) we find
\begin{equation}
[\mathcal{R}_{nloc},V_{b}]=\int\frac{d\mathbf{k}'}{\omega_{\mathbf{k}'}}
\int\frac{d\mathbf{k}}{\omega_{\mathbf{k}}}F_{b}^{\dag}[R(k'),G(k)]F_{b}a(k')a(k)
 \label{4.172}
\end{equation}
and
\[
[\mathcal{R}_{nloc},V_{b}^{\dag}]=\int\frac{d\mathbf{k}'}{\omega_{\mathbf{k}'}}
\int\frac{d\mathbf{k}}{\omega_{\mathbf{k}}}
\{F_{b}^{\dag}[R(k'),G(k_{-})]F_{b}a^{\dag}(k)a(k')
\]
\begin{equation}
+\delta(\mathbf{k'}-\mathbf{k}):F_{b}^{\dag}R(k')F_{b}::F_{b}^{\dag}G(k_{-})F_{b}:\},
 \label{4.173}
\end{equation}
where the matrix $G(k)$ is determined by Eq. (\ref{3.121}) and it
is implied that
\begin{equation}
[R(k'),G(k)](p',p)=\int\frac{d\mathbf{q}}{E_{\mathbf{q}}}
[R(p',q,k')G(q,p,k)-G(p',q,k)R(q,p,k')].
 \label{4.174}
\end{equation}

After the normal ordering of meson and boson operators in
commutator $[R_{nloc},V_{nloc}]$ one can obtain the
$2\rightarrow2$ interactions of the type $b^{\dag}a^{\dag}ba$,
$d^{\dag}a^{\dag}da$, $b^{\dag}d^{\dag}aa$, $a^{\dag}a^{\dag}bd$
and $b^{\dag}b^{\dag}bb$, $b^{\dag}d^{\dag}bd$,
$d^{\dag}d^{\dag}dd$ in the r.h.s. of Eqs. (\ref{4.172}) and
(\ref{4.173})and their H.c..

For example, the boson-boson interaction operator can be
represented as
\[
\frac{1}{2}[R_{nloc},V_{nloc}](bb\rightarrow
bb)=-\frac{1}{4}\int\frac{d\mathbf{p}'_{2}}{E_{\mathbf{p}'_{2}}}
\int\frac{d\mathbf{p}_{2}}{E_{\mathbf{p}_{2}}}\int\frac{d\mathbf{p}'_{1}}{E_{\mathbf{p}'_{1}}}
\int\frac{d\mathbf{p}_{1}}{E_{\mathbf{p}_{1}}}
\delta(\mathbf{p}'_{1}+\mathbf{p}'_{2}-\mathbf{p}_{1}-\mathbf{p}_{2})
\]
\[
\times g_{11}(p_{1}',p_{1},k)g_{11}(p_{2}',p_{2},k)
\]
\begin{equation}
\times
\left\{\frac{1}{(p_{1}-p_{1}')^{2}-\mu_{s}^{2}}+\frac{1}{(p_{2}-p_{2}')^{2}-\mu_{s}^{2}}\right\}
b^{\dag}_{c}(p'_{2})b^{\dag}_{c}(p_{1}')b_{c}(p_{2})b_{c}(p_{1})
 \label{4.175}
 \end{equation}
with $\mathbf{k}=\mathbf{p}'_{1}-\mathbf{p}_{1}$. Simultaneously,
we get the pair-production interaction operator
\[
\frac{1}{2}[R_{nloc},V_{nloc}](aa\rightarrow
b\bar{b})=\frac{1}{2}\int\frac{d\mathbf{p}'}{E_{\mathbf{p}'}}
\int\frac{d\mathbf{p}}{E_{\mathbf{p}}}\int\frac{d\mathbf{k}'}{\omega_{\mathbf{k}'}}
\int\frac{d\mathbf{k}}{\omega_{\mathbf{k}}}
\delta(\mathbf{p}'+\mathbf{p}-\mathbf{k}'-\mathbf{k})
\]
\[
\times (\frac{1}{E_{\mathbf{q}'}}
\frac{g_{11}(p',q',k')g_{12}(p,q',k)}
{E_{\mathbf{p}'}-E_{\mathbf{q}'}-\omega_{\mathbf{k}'}}
-\frac{1}{E_{\mathbf{q}'}}
\frac{g_{11}(p,q'_{-},k)g_{12}(p',q'_{-},k')}
{E_{\mathbf{p}'}+E_{\mathbf{q}'}-\omega_{\mathbf{k}'}}
\]
\[
+\frac{1}{E_{\mathbf{q}}} \frac{g_{12}(p',q',k')g_{11}(p,q',k)}
{E_{\mathbf{p}}-E_{\mathbf{q}}-\omega_{\mathbf{k}}})
-\frac{1}{E_{\mathbf{q}}}
\frac{g_{11}(p',q'_{-},k')g_{12}(p,q'_{-},k)}
{E_{\mathbf{p}}+E_{\mathbf{q}}-\omega_{\mathbf{k}}}
\]
\begin{equation}
\times b^{\dag}_{c}(p')d^{\dag}_{c}(p)a_{c}(k')a_{c}(k),
\label{4.176}
\end{equation}
where $\mathbf{q}'=\mathbf{p}'-\mathbf{k}'$,
$\mathbf{q}=\mathbf{p}-\mathbf{k}$ with the 4-momenta
$q'=(E_{\mathbf{q}'},\mathbf{q}')$ and
$q=(E_{\mathbf{q}},\mathbf{q})$.

In parallel, taking into account that in our model with
$\mathbf{N}_{bad}=\mathbf{N}_{B}$ we find the respective
contributions to $\mathbf{B}_{I}$,
\[
\frac{1}{2}[R_{nloc},\mathbf{N}_{B}](bb\rightarrow bb)
\]
\[
=\frac{i}{4}\int\frac{d\mathbf{p}'_{2}}{E_{\mathbf{p}'_{2}}}
\int\frac{d\mathbf{p}_{2}}{E_{\mathbf{p}_{2}}}\int\frac{d\mathbf{p}'_{1}}{E_{\mathbf{p}'_{1}}}
\int\frac{d\mathbf{p}_{1}}{E_{\mathbf{p}_{1}}}
\frac{\partial}{\partial\mathbf{p}'_{1}}\delta(\mathbf{p}'_{1}+\mathbf{p}'_{2}-\mathbf{p}_{1}-\mathbf{p}_{2})
\]
\[
\times g_{11}(p_{1}',p_{1},k)g_{11}(p_{2}',p_{2},k)
\]
\begin{equation}
\times
\left\{\frac{1}{(p_{1}-p_{1}')^{2}-\mu_{s}^{2}}+\frac{1}{(p_{2}-p_{2}')^{2}-\mu_{s}^{2}}\right\}
b^{\dag}_{c}(p'_{2})b^{\dag}_{c}(p_{1}')b_{c}(p_{2})b_{c}(p_{1})
 \label{4.177}
 \end{equation}
 and
\[
\frac{1}{2}[R_{nloc},\mathbf{N}_{B}](aa\rightarrow b\bar{b})
\]
\[
=-\frac{i}{2}\int\frac{d\mathbf{p}'}{E_{\mathbf{p}'}}
\int\frac{d\mathbf{p}}{E_{\mathbf{p}}}\int\frac{d\mathbf{k}'}{\omega_{\mathbf{k}'}}
\int\frac{d\mathbf{k}}{\omega_{\mathbf{k}}}
\frac{\partial}{\partial\mathbf{p}}\delta(\mathbf{p}+\mathbf{p}'-\mathbf{k}'-\mathbf{k})
\]
\[
\times (\frac{1}{E_{\mathbf{q}'}}
\frac{g_{11}(p',q',k')g_{12}(p,q',k)}
{E_{\mathbf{p}'}-E_{\mathbf{q}'}-\omega_{\mathbf{k}'}}
-\frac{1}{E_{\mathbf{q}'}}
\frac{g_{11}(p,q'_{-},k)g_{12}(p',q'_{-},k')}
{E_{\mathbf{p}'}+E_{\mathbf{q}'}-\omega_{\mathbf{k}'}}
\]
\[
+\frac{1}{E_{\mathbf{q}}} \frac{g_{12}(p',q',k')g_{11}(p,q',k)}
{E_{\mathbf{p}}-E_{\mathbf{q}}-\omega_{\mathbf{k}}}
-\frac{1}{E_{\mathbf{q}}}
\frac{g_{11}(p',q'_{-},k')g_{12}(p,q'_{-},k)}
{E_{\mathbf{p}}+E_{\mathbf{q}}-\omega_{\mathbf{k}}})
\]
\begin{equation}
\times b^{\dag}_{c}(p')d^{\dag}_{c}(p)a_{c}(k')a_{c}(k).
\label{4.178}
\end{equation}
In Eqs. (\ref{4.175}) and (\ref{4.177}) we meet a covariant
(Feynman-like) "propagator"
\begin{equation}
\frac{1}{2}\left\{\frac{1}{(p_{1}-p_{1}')^{2}-\mu_{s}^{2}}+\frac{1}{(p_{2}-p_{2}')^{2}-\mu_{s}^{2}}\right\},
\label{4.179}
 \end{equation}
which on the energy shell
\begin{equation}
E_{\mathbf{p}_{1}}+E_{\mathbf{p}_{1}}=E_{\mathbf{p}'_{1}}+E_{\mathbf{p}'_{2}}
\label{4.180}
 \end{equation}
is converted into the genuine Feynman propagator for the
corresponding S matrix (cf. discussions in \cite{SheShi01},
\cite{Korc93}).

\subsection{Mass renormalization and relativistic invariance}

We have seen how in the framework of the nonlocal meson-boson
model one can build the $2\rightarrow2$ interactions between the
clothed mesons and bosons. They appear in a natural way from the
commutator $\frac{1}{2}[R_{nloc},V_{nloc}]$ as the operators
$b^{\dag}a^{\dag}ba$, $d^{\dag}a^{\dag}da$, $b^{\dag}b^{\dag}bb$,
$b^{\dag}d^{\dag}bd$, $d^{\dag}d^{\dag}dd$, $b^{\dag}d^{\dag}aa$,
$a^{\dag}a^{\dag}bd$ of the class $[2.2]$. Moreover, this
commutator is a spring of the good operators $a^{\dag}a$,
$b^{\dag}b$ and $d^{\dag}d$ of the class $[1.1]$ together with the
bad operators $aa$ and $bd$ of the class
$[0.2]$\footnote{Henceforth, for brevity, we omit the subscript c}
and their hermitian conjugates $a^{\dag}a^{\dag}$ and
$b^{\dag}d^{\dag}$ of the class $[2.0]$. These operators may be
cancelled by the respective counterterms from
\begin{equation}
H_{nsc}(\alpha)=M_{s}(\alpha)+M_{b}(\alpha)
 \label{4.181}
 \end{equation}
in the r.h.s. of Eq. (\ref{4.159}). Let us show that such a
cancellation gives rise to certain definitions of the mass
coefficients in Eqs. (\ref{3.135}) and (\ref{3.136}).

Indeed, with the help of the same technique as in \cite{SheShi01}
one can show
\[
\frac{1}{2}[R_{nloc},V_{nloc}](a^{\dag}a)=-\frac{1}{2}\int\frac{d\mathbf{k}}{\omega_{\mathbf{k}}^{2}}
\int\frac{d\mathbf{p}}{E_{\mathbf{p}}E_{\mathbf{p}-\mathbf{k}}}[
\frac{g_{21}^{2}(p,q_{-},k_{-})}
{E_{\mathbf{p}}+E_{\mathbf{p}-\mathbf{k}}+\omega_{\mathbf{k}}}
\]
\begin{equation}
+\frac{g_{12}^{2}(p,q_{-},k)}
{E_{\mathbf{p}}+E_{\mathbf{p}-\mathbf{k}}-\omega_{\mathbf{k}}}
]a^{\dag}(k)a(k), \label{4.182}
\end{equation}
where $q=(E_{\mathbf{p}-\mathbf{k}},\mathbf{p}-\mathbf{k})$. In
the same way we obtain
\[
\frac{1}{2}[R_{nloc},V_{nloc}](aa)=-\frac{1}{2}\int\frac{d\mathbf{k}}{\omega_{\mathbf{k}}^{2}}
\int\frac{d\mathbf{p}}{E_{\mathbf{p}}E_{\mathbf{p}-\mathbf{k}}}
g_{12}(p,q_{-},k)g_{21}(p,q_{-},k_{-})
\]
\begin{equation}
\times[\frac{1}
{E_{\mathbf{p}}+E_{\mathbf{p}-\mathbf{k}}+\omega_{\mathbf{k}}}
+\frac{1}
{E_{\mathbf{p}}+E_{\mathbf{p}-\mathbf{k}}-\omega_{\mathbf{k}}}]a(k)a(k_{-})
\label{4.183}
\end{equation}
or
\[
\frac{1}{2}[R_{nloc},V_{nloc}](aa)=\int\frac{d\mathbf{k}}{\omega_{\mathbf{k}}^{2}}
\int\frac{d\mathbf{p}}{E_{\mathbf{p}}}
g_{12}(p,q_{-},k)g_{21}(p,q_{-},k_{-})
\]
\begin{equation}
\times[\frac{1}{\mu_{s}^{2}+2p_{-}k} +\frac{1}{\mu_{s}^{2}-2pk}]
a(k)a(k_{-}).
 \label{4.184}
\end{equation}
Recall that the last transition can be done by means of some trick
considered in Appendix A from \cite{SheShi01}.

Furthermore, assuming that
\begin{equation}
M_{s}^{(2)}(\alpha)+\frac{1}{2}[R_{nloc},V_{nloc}]_{2mes}=0
 \label{4.185}
 \end{equation}
with
\[
[R_{nloc},V_{nloc}]_{2mes}
\]
\[
=[R_{nloc},V_{nloc}](a^{\dag}a)+[R_{nloc},V_{nloc}](aa)
+[R_{nloc},V_{nloc}](a^{\dag}a^{\dag}),
\]
we find
\begin{equation}
m_{1}^{(2)}(k)=\frac{1}{2}\int\frac{d\mathbf{p}}{E_{\mathbf{p}}E_{\mathbf{p}-\mathbf{k}}}
[\frac{g_{21}^{2}(p,q_{-},k_{-})}
{E_{\mathbf{p}}+E_{\mathbf{p}-\mathbf{k}}+\omega_{\mathbf{k}}}+
 \frac{g_{12}^{2}(p,q_{-},k)}
 {E_{\mathbf{p}}+E_{\mathbf{p}-\mathbf{k}}-\omega_{\mathbf{k}}}]
\label{4.186}
 \end{equation}
and
\[
m_{2}^{(2)}(k)=-\int\frac{d\mathbf{p}}{E_{\mathbf{p}}}
g_{12}(p,q_{-},k)g_{21}(p,q_{-},k_{-})
\]
\begin{equation}
\times[\frac{1}{\mu_{s}^{2}+2p_{-}k} +\frac{1}{\mu_{s}^{2}-2pk}].
 \label{4.187}
 \end{equation}

The operators that conserve the boson (antiboson) number can be
written as (details see in \cite{KorShe04}):
\[
\frac{1}{2}[R_{nloc},V_{nloc}](b^{\dag}b)=\int\frac{d\mathbf{k}}{\omega_{\mathbf{k}}}
\int\frac{d\mathbf{p}}{E_{\mathbf{p}}^{2}E_{\mathbf{p}-\mathbf{k}}}
[\frac{g_{11}^{2}(p,q,k)}
{E_{\mathbf{p}}-E_{\mathbf{p}-\mathbf{k}}-\omega_{\mathbf{k}}}
\]
\begin{equation}
-\frac{g_{21}^{2}(p,q_{-},k_{-})}
{E_{\mathbf{p}}+E_{\mathbf{p}-\mathbf{k}}+\omega_{\mathbf{k}}}]b^{\dag}(p)b(p),
\label{4.188}
\end{equation}
\[
\frac{1}{2}[R_{nloc},V_{nloc}](d^{\dag}d)=\int\frac{d\mathbf{k}}{\omega_{\mathbf{k}}}
\int\frac{d\mathbf{p}}{E_{\mathbf{p}}^{2}E_{\mathbf{p}-\mathbf{k}}}
[\frac{g_{22}^{2}(p,q,k)}
{E_{\mathbf{p}}-E_{\mathbf{p}-\mathbf{k}}-\omega_{\mathbf{k}}}
\]
\begin{equation}
-\frac{g_{21}^{2}(p,q_{-},k_{-})}
{E_{\mathbf{p}}+E_{\mathbf{p}-\mathbf{k}}+\omega_{\mathbf{k}}}]d^{\dag}(p)d(p).
\label{4.189}
\end{equation}

One can show that from the condition
\begin{equation}
M_{b}^{(2)}(\alpha)+\frac{1}{2}[R_{nloc},V_{nloc}]_{2bos}=0,
 \label{4.190}
 \end{equation}
where
\[
[R_{nloc},V_{nloc}]_{2bos}=[R_{nloc},V_{nloc}](b^{\dag}b)+[R_{nloc},V_{nloc}](b^{\dag}d^{\dag})
\]
\[
+[R_{nloc},V_{nloc}](db)+[R_{nloc},V_{nloc}](d^{\dag}d)
\]
it follows
\begin{equation}
m_{11}^{(2)}(p)=-\int\frac{d\mathbf{k}}{\omega_{\mathbf{k}}E_{\mathbf{p}-\mathbf{k}}}
[\frac{g_{11}^{2}(p,q,k)}
{E_{\mathbf{p}}-E_{\mathbf{p}-\mathbf{k}}-\omega_{\mathbf{k}}}-
 \frac{g_{21}^{2}(p,q_{-},k_{-})}
 {E_{\mathbf{p}}+E_{\mathbf{p}-\mathbf{k}}+\omega_{\mathbf{k}}}],
\label{4.191}
 \end{equation}
\begin{equation}
m_{22}^{(2)}(p)=-\int\frac{d\mathbf{k}}{\omega_{\mathbf{k}}E_{\mathbf{p}-\mathbf{k}}}
[\frac{g_{11}^{2}(p,q,k)}
{E_{\mathbf{p}}-E_{\mathbf{p}-\mathbf{k}}-\omega_{\mathbf{k}}}-
 \frac{g_{21}^{2}(p,q_{-},k_{-})}
 {E_{\mathbf{p}}+E_{\mathbf{p}-\mathbf{k}}+\omega_{\mathbf{k}}}].
 \label{4.192}
 \end{equation}
Similarly one can obtain the non-diagonal coefficients
\[
m_{12}^{(2)}(p)=m_{21}^{(2)}(p)=-\int\frac{d\mathbf{k}}{\omega_{\mathbf{k}}E_{\mathbf{p}-\mathbf{k}}}
g_{11}(p,q,k)g_{21}(p,q_{-},k_{-})
\]
\begin{equation}
\times[\frac{1}{E_{\mathbf{p}}-E_{\mathbf{p}-\mathbf{k}}-\omega_{\mathbf{k}}}-
 \frac{1}{E_{\mathbf{p}}+E_{\mathbf{p}-\mathbf{k}}+\omega_{\mathbf{k}}}]
 \label{4.193}
 \end{equation}
or
\[
m_{12}^{(2)}(p)=m_{21}^{(2)}(p)
\]
\[
=-\int\frac{d\mathbf{k}}{\omega_{\mathbf{k}}}
g_{11}(p,q,k)g_{21}(p,q_{-},k_{-})[\frac{1}{\mu_{s}^{2}-2pk}+\frac{1}{\mu_{s}^{2}+2p_{-}k}]
\]
\begin{equation}
-\int\frac{d\mathbf{q}}{E_{\mathbf{q}}}
g_{11}(p,q,u)g_{21}(p,q_{-},u_{-})(\frac{1}{2[\mu_{b}^{2}-pq]-\mu_{s}^{2}}+\frac{1}{2[\mu_{b}^{2}+pq_{-}]-\mu_{s}^{2}}),
\label{4.194}
 \end{equation}
 where $u=(E_{\mathbf{p}-\mathbf{q}},\mathbf{p}-\mathbf{q})$.

The integrands in Eqs. (\ref{4.187}) and (\ref{4.194}) are
contained the  covariant denominators that have already occurred
in \cite{KorShe04} and \cite{SheShi01}. Thus the clothing
procedure has allowed us to get analytical expressions for the
interaction operators between the clothed particles. Moreover, we
have obtained some prescriptions when finding the coefficients in
the "mass renormalization" operators.

Unlike the momentum-independent mass shifts obtained in
\cite{KorShe04,SheShi01} and \cite{KruegGloe99} these
coefficients, as mentioned below Eq. (\ref{3.136}), may be
momentum dependent. But the most significant property of the
integrals (\ref{4.186}), (\ref{4.187}) and
(\ref{4.191})-(\ref{4.193}) is to take on finite values. In the
context, those divergent integrals from \cite{KorShe04,SheShi01},
being coincident with the Feynman one-loop ones for the pion and
nucleon mass shifts, are of interest as a prelude to the present
exploration.

At last, one should emphasize that if one starts from expansion
(\ref{3.72}) with the second-order contribution $H^{(2)}_{nsc}=0$,
then the RI would be violated at the beginning because of the
obvious discrepancy between Eqs. (\ref{3.74}) and (\ref{3.77}).

\section{Discussion. Towards working formulae}

We see that the way I in combination with the UCTs method makes
our consideration more and more appropriate for practical
applications (in particular, as one has to work with the vertex
cutoffs). It is well known that the role of such cutoffs may be
twofold, viz., first, as mentioned in Introduction to get rid of
ultraviolet divergences in the course of all intermediate
calculations and, second, to introduce the particle finite-size
effects. In this context, we will proceed with the $g$-factors,
which allow us, on the one hand, to do comparatively simple
calculations and, on the other hand, to preserve the basic
premises. In addition, of interest are their properties that could
provide the momentum independence of the particle mass shifts.

\subsection{The leading-order mass shifts and their momentum dependence}

The formulae for the $2\rightarrow2$ interactions in Subsec. 4.1
and for the mass coefficients in Subsec. 4.2 become more tractable
if we assume that
\begin{equation}
g_{\varepsilon'\varepsilon}(p',p,k)=v_{\varepsilon'\varepsilon}
([k+(-1)^{\varepsilon'}p'-(-1)^{\varepsilon}p][k-(-1)^{\varepsilon'}p'+(-1)^{\varepsilon}p]).
\label{5.195}
 \end{equation}
One can verify the nonlocal model with such cutoffs possesses
necessary properties (\ref{3.128})-(\ref{3.131}). In terms of the
$v_{\varepsilon'\varepsilon}$ functions we get
\[
m_{1}^{(2)}(k)=\frac{1}{2}\int\frac{d\mathbf{p}}{E_{\mathbf{p}}E_{\mathbf{p}-\mathbf{k}}}
[\frac{v_{21}^{2}(\omega_{\mathbf{k}}^{2}-(E_{\mathbf{p}}+E_{\mathbf{p}-\mathbf{k}})^{2})}
{E_{\mathbf{p}}+E_{\mathbf{p}-\mathbf{k}}+\omega_{\mathbf{k}}}
\]
\begin{equation}
+\frac{v_{12}^{2}(\omega_{\mathbf{k}}^{2}-(E_{\mathbf{p}}+E_{\mathbf{p}-\mathbf{k}})^{2})}
 {E_{\mathbf{p}}+E_{\mathbf{p}-\mathbf{k}}-\omega_{\mathbf{k}}}]
\label{5.196}
 \end{equation}
and
\[
m_{2}^{(2)}(k)=-\int\frac{d\mathbf{p}}{E_{\mathbf{p}}}
v_{21}(\omega_{\mathbf{k}}^{2}-(E_{\mathbf{p}}+E_{\mathbf{p}-\mathbf{k}})^{2})
v_{12}(\omega_{\mathbf{k}}^{2}-(E_{\mathbf{p}}+E_{\mathbf{p}-\mathbf{k}})^{2})
\]
\begin{equation}
\times[\frac{1}{\mu_{s}^{2}+2p_{-}k} +\frac{1}{\mu_{s}^{2}-2pk}].
 \label{5.197}
 \end{equation}
 Now, by handling the charge-independent cutoffs,
\begin{equation}
v_{12}(x)=v_{21}(x)=f(x), \label{5.198}
 \end{equation}
 we obtain
\[
m_{1}^{(2)}(k)=m_{2}^{(2)}(k)
\]
\[
=\int\frac{d\mathbf{p}}{E_{\mathbf{p}}E_{\mathbf{p}-\mathbf{k}}}
(E_{\mathbf{p}}+E_{\mathbf{p}-\mathbf{k}})
\frac{f^{2}(\omega_{\mathbf{k}}^{2}-(E_{\mathbf{p}}+E_{\mathbf{p}-\mathbf{k}})^{2})}
{(E_{\mathbf{p}}+E_{\mathbf{p}-\mathbf{k}})^{2}-\omega_{\mathbf{k}}^{2}}
\]
\[
=-\int\frac{d\mathbf{p}}{E_{\mathbf{p}}}
f^{2}(\omega_{\mathbf{k}}^{2}-(E_{\mathbf{p}}+E_{\mathbf{p}-\mathbf{k}})^{2})
[\frac{1}{\mu_{s}^{2}+2p_{-}k} +\frac{1}{\mu_{s}^{2}-2pk}]
\]
\begin{equation}
=-\int\frac{d\mathbf{p}}{E_{\mathbf{p}}}
\frac{f^{2}(\omega_{\mathbf{k}}^{2}-(E_{\mathbf{p}}+E_{\mathbf{p}+\mathbf{k}})^{2})}
{{\mu_{s}^{2}+2pk}}- \int\frac{d\mathbf{p}}{E_{\mathbf{p}}}
\frac{f^{2}(\omega_{\mathbf{k}}^{2}-(E_{\mathbf{p}}+E_{\mathbf{p}-\mathbf{k}})^{2})}
{{\mu_{s}^{2}-2pk}}. \label{5.199}
\end{equation}
The second form of these coefficients has been prompted by the
trick \cite{SheShi01} with
\[
\frac{E_{\mathbf{p}}+E_{\mathbf{p}-\mathbf{k}}}
{(E_{\mathbf{p}}+E_{\mathbf{p}-\mathbf{k}})^{2}-\omega_{\mathbf{k}}^{2}}=
-E_{\mathbf{p}-\mathbf{k}}(\frac{1}{\mu_{s}^{2}-2pk}+\frac{1}{\mu_{s}^{2}+2p_{-}k})
+\frac{E_{\mathbf{p}}-E_{\mathbf{p}-\mathbf{k}}}
{(E_{\mathbf{p}}-E_{\mathbf{p}-\mathbf{k}})^{2}-\omega_{\mathbf{k}}^{2}}
\]
and using the properties (\ref{3.129})-(\ref{3.131}). In other
words, the option (\ref{5.198}) yields the momentum-independent
coefficients $m_{1}^{(2)}(k)=m_{2}^{(2)}(k)\equiv m_{s}^{(2)}$.
Indeed, along with the Lorentz invariant denominators the
integrand in the r.h.s. of (\ref{5.199}) is contained function
$f(I)$ whose argument
\[
I(\mathbf{p},\mathbf{k})\equiv\omega_{\mathbf{k}}^{2}-(E_{\mathbf{p}}+E_{\mathbf{p}-\mathbf{k}})^{2}
=\mu_{s}^{2}-2\mu_{b}^{2}-2E_{\mathbf{p}}E_{\mathbf{p}-\mathbf{k}}-2\mathbf{p}(\mathbf{p}-\mathbf{k})
\]
does not change under the simultaneous transformation
$\mathbf{p}\Rightarrow\mathbf{p'}=\mathbf{\Lambda p}$ and
$\mathbf{p}-\mathbf{k}\Rightarrow\mathbf{\Lambda (p-k)}$ on the
mass shells $p^{2}=\mu_{b}^{2}$ and $k^{2}=\mu_{s}^{2}$. Similar
combinations have been considered in \cite{Shi2000}, where the
author, handling the mass renormalization problem within
noncovariant perturbation theory for a nonlocal extension of the
Wentzel model, gives some reasonings in favor of the momentum
independence of such integrals as (\ref{5.199}). In particular, he
has addressed earlier works \cite{KawabeUmezava1,KawabeUmezava2}
in which similar evaluations have been carried out by means of a
cumbersome procedure with so-called $w$-transformation of
integration variables. By invoking those results, one can reduce
the triple integral to the simple one,
\begin{equation}
m_{s}^{(2)}=8\pi\int_{0}^{\infty}\frac{t^{2}dt}{\sqrt{t^{2}+\mu_{b}^{2}}}\frac{f^{2}(\mu_{s}^{2}-4t^{2}-4\mu_{b}^{2})}{4t^{2}+4\mu_{b}^{2}-\mu_{s}^{2}}.
\label{5.200}
\end{equation}
For our purposes it suffices to use alternate derivation of this
result, given in Appendix C.

Furthermore, from Eqs. (\ref{4.191})-(\ref{4.193}) it follows
\[
 m_{11}^{(2)}(p)=m_{22}^{(2)}(p)
\]
\begin{equation}
=-\int\frac{d\mathbf{k}}{\omega_{\mathbf{k}}E_{\mathbf{p}-\mathbf{k}}}
[\frac{v_{11}^{2}(\omega_{\mathbf{k}}^{2}-(E_{\mathbf{p}}-E_{\mathbf{p}-\mathbf{k}})^{2})}
{E_{\mathbf{p}}-E_{\mathbf{p}-\mathbf{k}}-\omega_{\mathbf{k}}}
-\frac{v_{21}^{2}(\omega_{\mathbf{k}}^{2}-(E_{\mathbf{p}}+E_{\mathbf{p}-\mathbf{k}})^{2})}
 {E_{\mathbf{p}}+E_{\mathbf{p}-\mathbf{k}}+\omega_{\mathbf{k}}}],
\label{5.201}
 \end{equation}
\[
 m_{12}^{(2)}(p)=m_{21}^{(2)}(p)
\]
\[
=-\int\frac{d\mathbf{k}}{\omega_{\mathbf{k}}E_{\mathbf{p}-\mathbf{k}}}
v_{11}(\omega_{\mathbf{k}}^{2}-(E_{\mathbf{p}}-E_{\mathbf{p}-\mathbf{k}})^{2})
v_{21}(\omega_{\mathbf{k}}^{2}-(E_{\mathbf{p}}+E_{\mathbf{p}-\mathbf{k}})^{2})
\]
\begin{equation}
\times[\frac{1}{E_{\mathbf{p}}-E_{\mathbf{p}-\mathbf{k}}-\omega_{\mathbf{k}}}-
 \frac{1}{E_{\mathbf{p}}+E_{\mathbf{p}-\mathbf{k}}+\omega_{\mathbf{k}}}].
 \label{5.202}
 \end{equation}

Evaluation of these coefficients is simplified once we put
\[
v_{11}(\omega_{\mathbf{k}}^{2}-(E_{\mathbf{p}}-E_{\mathbf{p}-\mathbf{k}})^{2})=
 v_{21}(\omega_{\mathbf{k}}^{2}-(E_{\mathbf{p}}+E_{\mathbf{p}-\mathbf{k}})^{2})
\]
\begin{equation}
 =f(\omega_{\mathbf{k}}^{2}-(E_{\mathbf{p}}+E_{\mathbf{p}-\mathbf{k}})^{2}).
\label{5.203}
 \end{equation}
\[
m_{b}^{(2)}(p)\equiv
m_{11}^{(2)}(p)=m_{21}^{(2)}(p)=-\int\frac{d\mathbf{k}}{\omega_{\mathbf{k}}E_{\mathbf{p}-\mathbf{k}}}
f^{2}(\omega_{\mathbf{k}}^{2}-(E_{\mathbf{p}}+E_{\mathbf{p}-\mathbf{k}})^{2})
\]
\[
\times[\frac{1}{E_{\mathbf{p}}-E_{\mathbf{p}-\mathbf{k}}-\omega_{\mathbf{k}}}-
 \frac{1}{E_{\mathbf{p}}+E_{\mathbf{p}-\mathbf{k}}+\omega_{\mathbf{k}}}]
\]
\begin{equation}
=2\int\frac{d\mathbf{k}}{\omega_{\mathbf{k}}}
\frac{f^{2}(\omega_{\mathbf{k}}^{2}-(E_{\mathbf{p}}+E_{\mathbf{p}-\mathbf{k}})^{2})}
{E_{\mathbf{p}-\mathbf{k}}^{2}-(E_{\mathbf{p}}-\omega_{\mathbf{k}})^{2}}
+2\int\frac{d\mathbf{k}}{E_{\mathbf{p}-\mathbf{k}}}
\frac{f^{2}(\omega_{\mathbf{k}}^{2}-(E_{\mathbf{p}}+E_{\mathbf{p}-\mathbf{k}})^{2})}
{\omega_{\mathbf{k}}^{2}-(E_{\mathbf{p}}+E_{\mathbf{p}-\mathbf{k}})^{2}}
 \label{5.204}
\end{equation}
or
\[
m_{b}^{(2)}(p)=C_{1}(p)+C_{2}(p),
\]
\[
C_{1}(p)=2\int\frac{d\mathbf{k}}{\omega_{\mathbf{k}}}
\frac{f^{2}(\omega_{\mathbf{k}}^{2}-(E_{\mathbf{p}}+E_{\mathbf{p}-\mathbf{k}})^{2})}
{2pk-\mu_{s}^{2}}
\]
and
\[
C_{2}(p)=2\int\frac{d\mathbf{q}}{E_{\mathbf{q}}}
\frac{f^{2}(\mu_{s}^{2}-2\mu_{b}^{2}-2pq)}
{\mu_{s}^{2}-2\mu_{b}^{2}-2pq}.
\]
Evidently, the second integral does not depend upon $p$ so
\[
C_{2}(p)=C_{2}(0)=2\int\frac{d\mathbf{q}}{E_{\mathbf{q}}}
\frac{f^{2}(\mu_{s}^{2}-2\mu_{b}^{2}-2\mu_{b}E_{\mathbf{q}})}
{\mu_{s}^{2}-2\mu_{b}^{2}-2\mu_{b}E_{\mathbf{q}}}
\]
\begin{equation}
=8\pi\int_{0}^{\infty}\frac{q^{2}dq}{E_{\mathbf{q}}}
\frac{f^{2}(\mu_{s}^{2}-2\mu_{b}^{2}-2\mu_{b}E_{\mathbf{q}})}
{\mu_{s}^{2}-2\mu_{b}^{2}-2\mu_{b}E_{\mathbf{q}}}. \label{5.205}
\end{equation}
It is not the case for integral $C_{1}(p)$. Thus under the link
(\ref{5.203}) the boson "mass renormalization" coefficients may be
momentum dependent (cf. our comment below Eq. (\ref{3.136})).

At the point one should realize that within our approach, where we
are trying to do without any fantom such as the bare masses and
coupling constants, the one-meson and one-boson operators $M_{s}$
and $M_{b}$ cannot appear in the new form $K(\alpha_{c})$ of the
initial Hamiltonian. Their main destination is to provide the RI
as whole (see the note below Eq. (\ref{3.148})) and we  have seen
how the second-order displacement $\mathbf{D}^{(2)}$ by Eqs.
(\ref{3.149})-(\ref{3.151}) and higher-order contributions to the
boost operator can be evaluated in the CPR. It is important that
the integrals $m_{s}^{(2)}$, $C_{1}(p)$ and $C_{2}(p)$ are
convergent at proper choice of the cutoff function. Moreover, as
shown in Appendix C, the $m_{s}^{(2)}$ value considerably
decreases when moving from the large $\Lambda$ values (smeared
cutoffs) to smaller $\Lambda$'s, i.e., cutoffs more localized in
momentum space. It is equivalent to an effective weakening of the
initial nonlocal interaction with its coupling constant $g$.
A similar trend takes place for other "renormalization" integrals
$C_{1}(p)$ and $C_{2}(0)$ when the former changes very slowly with
the $p$ increase starting from $p$ values comparable to a fixed
$\Lambda$. These results give us a spring of inspiration for
future explorations of the convergence of the recursive procedure
proposed here.

Of course, the introduction of a unique cutoff factor $f(x)$
simplifies the interpretation of the integrals obtained in Subsec.
4.1. In fact, under the conditions (\ref{5.198}) and (\ref{5.203})
we find, for example,
\[
\frac{1}{2}[R_{nloc},V_{nloc}](aa\rightarrow
b\bar{b})=\int\frac{d\mathbf{p}'}{E_{\mathbf{p}'}}
\int\frac{d\mathbf{p}}{E_{\mathbf{p}}}\int\frac{d\mathbf{k}}{\omega_{\mathbf{k}}}
\int\frac{d\mathbf{k}'}{\omega_{\mathbf{k}'}}
\delta(\mathbf{p}'+\mathbf{p}-\mathbf{k}'-\mathbf{k})
\]
\[
\times
f(\omega_{\mathbf{k'}}^{2}-(E_{\mathbf{p'}}+E_{\mathbf{p'}-\mathbf{k'}})^{2})
f(\omega_{\mathbf{k}}^{2}-(E_{\mathbf{p}}+E_{\mathbf{p}-\mathbf{k}})^{2})
\]
\begin{equation}
\times[\frac{1} {(p'-k')^{2}-\mu_{b}^{2}} +\frac{1}
{(p-k)^{2}-\mu_{b}^{2}}]
b^{\dag}_{c}(p')d^{\dag}_{c}(p)a_{c}(k')a_{c}(k). \label{5.206}
\end{equation}
Again, we encounter the Feynman-like "propagator", which on the
energy shell is converted into the true Feynman propagator for the
corresponding S matrix. Moreover, it turns out that the commutator
\[
[V_{nloc}(t,\mathbf{x'}),V_{nloc}(t,\mathbf{x})]=0
\]
under the constraints (\ref{5.198}) and (\ref{5.203}) too (cf. Eq.
(\ref{3.147})). Thus, we see that the correction
$\mathbf{D}^{(2)}$ is determined by
\begin{equation}
\mathbf{D}^{(2)}=\mathbf{D}_{con}^{(2)}+\mathbf{D}_{ncon}^{(2)},
 \label{5.207}
\end{equation}
\[
\mathbf{D}_{con}^{(2)}=\frac{i}{2}m_{s}^{(2)}
\int\frac{d\mathbf{k}}{\omega_{\mathbf{k}}^{2}} (\frac{\partial
a^{\dagger }(k)}{\partial
\mathbf{k}}a(k)-a^{\dagger}(k)\frac{\partial a(k)}{\partial
\mathbf{k}})
\]
\[
+\frac{i}{2}C_{2}(0)\int\frac{d\mathbf{p}}{E_{\mathbf{p}}^{2}}
(\frac{\partial b^{\dagger }(p)}{\partial
\mathbf{p}}b(p)-b^{\dagger}(p)\frac{\partial b(p)}{\partial
\mathbf{p}} +\frac{\partial d^{\dagger }(p)}{\partial
\mathbf{p}}d(p)-d^{\dagger}(p)\frac{\partial d(p)}{\partial
\mathbf{p}})
\]
\[
+\frac{i}{2}\int\frac{d\mathbf{p}'}{E_{\mathbf{p}'}}
\int\frac{d\mathbf{p}}{E_{\mathbf{p}}}(E_{\mathbf{p}'}E_{\mathbf{p}}+\mathbf{p}'\mathbf{p}+\mu_{b}^{2})
(\frac{C_{1}(p)}{E_{\mathbf{p}}}-\frac{C_{1}(p')}{E_{\mathbf{p}'}})
\]
\begin{equation}
\times\frac{b^{\dag}(p')b(p)+d^{\dag}(p')d(p)}
{E_{\mathbf{p}'}-E_{\mathbf{p}}}
\frac{\partial}{\partial\mathbf{p}}\delta(\mathbf{p}-\mathbf{p}')
\label{5.208}
\end{equation}
and
\[
\mathbf{D}_{ncon}^{(2)}= \frac{i}{2}m_{s}^{(2)}
\int\frac{d\mathbf{k}}{\omega_{\mathbf{k}}}\frac{\mathbf{k}}{\omega_{\mathbf{k}}^{3}}
(a^{\dagger }(k)a^{\dagger }(k_{-})-a(k)a(k_{-}))
\]
\[
+im_{s}^{(2)} \int\frac{d\mathbf{k}}{\omega_{\mathbf{k}}^{2}}
(\frac{\partial a^{\dagger }(k)}{\partial \mathbf{k}}a^{\dagger
}(k_{-})-a(k_{-})\frac{\partial a(k)}{\partial \mathbf{k}})
\]
\[
+\frac{i}{2}C_{2}(0)
\int\frac{d\mathbf{p}}{E_{\mathbf{p}}}\frac{\mathbf{p}}{E_{\mathbf{p}}^{3}}
(b^{\dagger }(p)d^{\dagger }(p_{-})-b(p)d(p_{-}))
\]
\[
+iC_{2}(0) \int\frac{d\mathbf{p}}{E_{\mathbf{p}}^{2}}
(\frac{\partial b^{\dagger }(p)}{\partial \mathbf{p}}d^{\dagger
}(p_{-})-b^{\dagger }(p)\frac{\partial d^{\dagger
}(p_{-})}{\partial \mathbf{p}}
 -\frac{\partial b(p)}{\partial
\mathbf{p}}d(p_{-})+b(p)\frac{\partial d(p_{-})}{\partial
\mathbf{p}})
\]
\[
+i\int\frac{d\mathbf{p}'}{E_{\mathbf{p}'}}
\int\frac{d\mathbf{p}}{E_{\mathbf{p}}}
C_{1}(p)\frac{E_{\mathbf{p}'}E_{\mathbf{p}}+\mathbf{p}'\mathbf{p}+\mu_{b}^{2}}{E_{\mathbf{p}}}
\]
\begin{equation}
\times\frac{b^{\dag}(p')d^{\dag}(p_{-})-b(p')d(p_{-})}{E_{\mathbf{p'}}+E_{\mathbf{p}}}
\frac{\partial}{\partial\mathbf{p}}\delta(\mathbf{p}-\mathbf{p}').
\label{5.209}
\end{equation}
Being compared with the free boosts by Eqs. (\ref{A.14}) and
(\ref{A.15}) (of course, both in the CPR) the correction
(\ref{5.207}) reduces to replacements of
\[
\frac{1}{\sqrt{\omega_{\mathbf{k}}\omega_{\mathbf{k}'}}}
\longrightarrow
\frac{1}{\sqrt{\omega_{\mathbf{k}}\omega_{\mathbf{k}'}}}
(1+\frac{m_{s}^{(2)}}{\omega_{\mathbf{k}}\omega_{\mathbf{k}'}})
\]
and
\[
\frac{1}{\sqrt{E_{\mathbf{p}}E_{\mathbf{p}'}}} \longrightarrow
\frac{1}{\sqrt{E_{\mathbf{p}}E_{\mathbf{p}'}}}
(1+\frac{C_{2}(0)}{E_{\mathbf{p}}E_{\mathbf{p}'}}+\frac{1}{E_{\mathbf{p}'}-E_{\mathbf{p}}}
[\frac{C_{1}(p)}{E_{\mathbf{p}}}-\frac{C_{1}(p')}{E_{\mathbf{p}'}}]),
\]
respectively, in the integrands for the meson boost and the boson
boost. It turns out that at moderate $\Lambda$ values $\sim $ 1 GeV (typical of the theory of meson-nucleon interactions)
in the
cutoff function (\ref{C.9}) the respective numerical deviations from the free boosts can be small.

\subsection{Deuteron properties in the CPR}
Besides, we would like to outline the basic elements of another
our exploration that is in progress. It is the case, where relying
upon the available experience of relativistic calculations of the
deuteron static moments in
\cite{ChungKeiCoester}-\cite{LevPaceSalme} and the deuteron FFs
(see reviews \cite{BonBurMolSmir}-\cite{GilmanGross} and refs.
therein) one has to deal with the matrix elements
$\langle\mathbf{P}',M'|J^{\mu}(0)|\mathbf{P}=0,M\rangle$ (to be
definite in the laboratory frame). Here the operator $J^{\mu}(0)$
is the N\"{o}ther current density $J^{\mu}(x)$ at $x=0$,
sandwiched between the eigenstates of a "strong" field Hamiltonian
$H$ (cf., discussion in Sec. 5 of lecture \cite{SheShi00}). In the
CPR with $H=K(\alpha_{c})$ (Eq. (\ref{4.158})) and
$\mathbf{N}=\mathbf{B}(\alpha_{c})$ (Eq. (\ref{4.160})) the
deuteron state $|\mathbf{P}=0,M\rangle$
($|\mathbf{P'}=\mathbf{q},M'\rangle$) in the rest (the frame
moving with the velocity $\mathbf{v}=\mathbf{q}/m_{d}$) meets the
eigenvalue equation
\begin{equation}
P^{\mu}|\mathbf{P},M\rangle=P_{d}^{\mu}|\mathbf{P},M\rangle
 \label{5.210}
\end{equation}
with the three-momentum transfer $\mathbf{q}$, four-momentum
$P_{d}^{\mu}=(E_{d},\mathbf{P})$,
$E_{d}=\sqrt{\mathbf{P}^{2}+m_{d}^{2}}$,
$m_{d}=m_{p}+m_{n}-\varepsilon_{d}$ and the deuteron binding
energy $\varepsilon_{d}>0$.

We know that such observables as the charge, magnetic and
quadrupole moments of the deuteron can be expressed through the
matrix elements in question (e.g., within the Bethe-Salpeter (BS)
formalism \cite{BonBurMolSmir}-\cite{GilmanGross}), where,
according to the original contribution \cite{Glas57}, one
introduces the corresponding covariant FFs. With the aid of
cumbersome numerical methods the latter have been evaluated in
terms of the Mandelstam current sandwiched between the deuteron BS
amplitudes. Some results in the subfield one can find in
\cite{KoShe88}, \cite{Burov96}.

Unlike this, following \cite{SheShi00} and \cite{DuFroShe10}, we
consider the expansion in the $R$-commutators
\begin{equation}
J^{\mu}(0)=WJ^{\mu}_{c}(0)W^{\dag}=J^{\mu}_{c}(0)+[R,J^{\mu}_{c}(0)]+\frac{1}{2}[R,[R,J^{\mu}_{c}(0)]]+...,
 \label{5.211}
\end{equation}
where $J^{\mu}_{c}(0)$ is the initial current in which the bare
operators $\{\alpha\}$ are replaced by the clothed ones
$\{\alpha_{c}\}$. Decomposition (\ref{5.211}) involves one-body,
two-body and more complicated interaction currents, if one uses
the terminology  customary in the theory of meson exchange
currents (MEC) \cite{MesInNucl}. Further, to the approximation
\begin{equation}
K_{I}=K(NN\rightarrow NN)\sim b_{c}^{\dag}b_{c}^{\dag}b_{c}b_{c}
 \label{5.212}
\end{equation}
and
\begin{equation}
\mathbf{B}_{I}=\mathbf{B}(NN\rightarrow NN)\sim
b_{c}^{\dag}b_{c}^{\dag}b_{c}b_{c}
 \label{5.213}
\end{equation}
(see, respectively, (\ref{4.175}) and (\ref{4.177})) the
eigenvalue problem (\ref{5.210}) becomes simpler so its solution
acquires the form
\begin{equation}
|\mathbf{P},M\rangle=\int d\mathbf{p}_{1}\int
d\mathbf{p}_{2}D_{M}([\mathbf{P}];\mathbf{p}_{1}\mu_{1};\mathbf{p}_{2}\mu_{2})
b_{c}^{\dag}(\mathbf{p}_{1}\mu_{1})b_{c}^{\dag}(\mathbf{p}_{2}\mu_{2})|\Omega\rangle.
 \label{5.214}
\end{equation}
In this connection, let us recall the relation
\begin{equation}
|\mathbf{q},M\rangle=exp[i\mathbf{\beta}\mathbf{B}(\alpha_{c})]|\mathbf{0},M\rangle
 \label{5.215}
\end{equation}
with $\mathbf{\beta}=\beta\mathbf{n}$, $\mathbf{n}=\mathbf{n}/n$
and $\tanh\beta=v$, that takes place owing to the property
\begin{equation}
e^{i\mathbf{\beta}\mathbf{B}}P^{\mu}e^{-i\mathbf{\beta}\mathbf{B}}=
P^{\nu}L_{\nu}^{\mu}(\mathbf{\beta}),
 \label{5.216}
\end{equation}
where $L(\mathbf{\beta})$ is the matrix of the corresponding
Lorentz transformation. Note also that the label $M=(\pm1,0)$
denotes the eigenvalue of the third component of the total (field)
angular-momentum operator in the deuteron center-of-mass (details
can be found in \cite{FBS2010}). The $c$-coefficients $D_{M}$ in
Eq. (\ref{5.214}) are calculated by solving the homogeneous
Lippmann-Schwinger equation with the quasipotentials taken from
\cite{FBS2010} (see formulae (67)-(69) therein). Numerical results
can be obtained either using the angular-momentum decomposition
(as in \cite{FBS2010}) or without it (as in \cite{HammeGlockle},
\cite{ElsterGlockle}). In other words, we are able to do without a
semirelativistic treatment, where only lowest order relativistic
contributions are included (see \cite{Arenh00} and refs. therein).

In its turn, the operator (\ref{5.211}) being between the clothed
two-nucleon states contributes as
\begin{equation}
\eta_{c}J^{\mu}(0)\eta_{c}=J_{one-body}^{\mu}+J_{two-body}^{\mu},
 \label{5.217}
\end{equation}
where the operator
\begin{equation}
J_{one-body}^{\mu}=\int
d\mathbf{p}'d\mathbf{p}F_{p,n}^{\mu}(\mathbf{p}',\mathbf{p})b_{c}^{\dag}(\mathbf{p})b_{c}(\mathbf{p})
 \label{5.218}
\end{equation}
with
\begin{equation}
F_{p,n}^{\mu}(\mathbf{p}',\mathbf{p})=e\bar{u}(\mathbf{p}'){F_{1}^{p,n}[(p'-p)^{2}]\gamma^{\mu}+
i\sigma^{\mu\nu}(p'-p)_{\nu}F_{2}^{p,n}[(p'-p)^{2}]}u(\mathbf{p})
 \label{5.219}
\end{equation}
that describes the virtual photon interaction with the clothed
proton (neutron)\footnote{In Eqs. (\ref{5.217}) $\eta_{c}$ is the
projection operator on the subspace $\mathcal{H}_{2N}$ $\in $
$\mathcal{H}$ spanned on the two-clothed-nucleon states
$|2N\rangle=b_{c}^{\dag}b_{c}^{\dag}|\Omega\rangle$}.

Its appearance follows from the observation, in which the primary
N\"{o}ther current operator, being between the physical (clothed)
states $|\Psi_{N}\rangle=b_{c}^{\dag}|\Omega\rangle$, yields the
usual on-mass-shell expression
\[
\langle\Psi_{p,n}(\mathbf{p}')|J^{\mu}(0)|\Psi_{p,n}(\mathbf{p})\rangle=
F_{p,n}^{\mu}(\mathbf{p}',\mathbf{p})
\]
in terms of the Dirac and Pauli nucleon FFs.\footnote{Of course,
all nucleon polarization labels are implied here together with
necessary summations over them in Eq. (\ref{5.218}) and so on}

By keeping in the r.h.s. of Eq. (\ref{5.217}) only the one-body
contribution we arrive to certain off-energy-shell extrapolation
of the so-called relativistic impulse approximation (RIA) in the
theory of e.m. interactions with nuclei (bound systems). In a
recent work by Dubovyk and Shebeko the deuteron magnetic and
quadrupole moments have been calculated, using the RIA, to be
submitted to Few Body Systems, where the previous paper
\cite{FBS2010} has been published.

Of course, the RIA results should be corrected including more
complex mechanisms of e-d scattering, that are contained in
\begin{equation}
J_{two-body}^{\mu}=\int
d\mathbf{p}_{1}'d\mathbf{p}_{2}'d\mathbf{p}_{1}d\mathbf{p}_{2}F_{MEC}^{\mu}
(\mathbf{p}_{1}',\mathbf{p}_{2}';\mathbf{p}_{1},\mathbf{p}_{2})
b_{c}^{\dag}(\mathbf{p}_{1}')b_{c}^{\dag}(\mathbf{p}_{1}')
b_{c}(\mathbf{p}_{1})b_{c}(\mathbf{p}_{2}).
 \label{5.220}
\end{equation}
Analytic (approximate) expressions for the coefficients
$F_{MEC}^{\mu}$ stem from the $R$-commutators (beginning with the
third one) in the expansion (\ref{5.211}), which, first, belong to
the class $[2.2]$, as in Eq. (\ref{5.218}), and, second, depend on
even numbers of mesons involved. It requires a separate
consideration aimed at finding a new family of MEC, as we hope not
only for the e-d scattering.

At last, one should note that, as before, we prefer to handle the
explicitly gauge-independent (GI) representation of photonuclear
reaction amplitudes with one-proton absorption or emittance
\cite{LevShe93}, \cite{CanLevShe2004}. This representation is an
extension of the Siegert theorem, in which, the amplitude of
interest is expressed through the Fourier transforms of electric
(magnetic) field strenghts and the generalized electric (magnetic)
dipole moments of hadronic system. It allows us to retain the GI
in the course of inevitably approximate calculations.

\section{Summary}
We propose a constructive way of ensuring the RI in QFT with
cutoffs in momentum space. In contrast to the traditional
approach, where the generators of $\Pi$ are determined as the
Noether integrals of the energy-momentum density tensor, we do not
utilize the Lagrangian formalism so fruitful in case of local
field models. Our purpose is to find these generators as elements
of the Lie algebra of $\Pi$ starting from the total Hamiltonian
whose interaction density in the Dirac picture includes a
Lorentz-scalar part $H_{sc}(x)$. Respectively, the algebraic
aspect of the RI as a whole for the present exploration with the
so-called instant form of relativistic dynamics is of paramount
importance.

In the context, using purely algebraic means the boost generators
can be decomposed into the Belinfante operator built of $H_{sc}$
and the operator which accumulates the chain of recursive
relations in the second and higher orders in $H_{nsc}$. Thereby,
it becomes clear that Poincar\'{e} commutations are not fulfilled
if the Hamiltonian does not contain some additional ingredients,
which we call the mass renormalization terms, though beyond local
field models such a terminology looks rather conventional. We have
shown how the method of UCTs enables us to determine the
corresponding operators for a given model. Moreover, it can be done using its nonlocal
extensions satisfying the requirements of special relativity and preserving certain continuity
with local QFTs.

We see that our approach is sufficiently flexible being applied
not merely to local field models including ones with derivative
couplings and spin $j\geq1$. Its realization, shown here for the
nonlocal extensions of the well-known Yukawa-type couplings, gives
us an encouraging impetus when constructing the interactions
between the clothed particles simultaneously in the Hamiltonian
and the corresponding boost operator. In the course of such a work
that is under way (see Subsec 5.2) we are trying to understand to
what extent the deuteron quenching in flight affects the deuteron
electromagnetic form factors. In our opinion, the present
exploration may be also helpful for a field-theoretical treatment
of particle decays in flight.

The RI of the S-matrix, that follows from the RI as a whole,  can be employed in future calculations, first, in the Dirac
picture owing to a unitary equivalence of the CPR to the BPR and, second,
in the Heisenberg picture after finding certain links between the
in (out) states and the clothed-particle ones (see our talk in
Durham \cite{She03}). It is known that the latter is most
appropriate for describing collisions with the bound systems. We
are ready to show our results in these directions somewhere else.

At last, we have tried to offer not only a fresh look at constructing
the generators in question but also a nonstandard renormalization
procedure in relativistic quantum field theory.
In this context, let us remind the prophetic words by
Dirac \cite {Dirac63}: "I am inclined to suspect that
renormalization theory is something that will not survive in the
future, and the remarkable agreement between theory and experiment
should be looked on as a fluke".

\appendix

\section{Generators of the Poincar\'{e} group in the BPR for free pion and nucleon fields}
Replacing in the free densities $H_{\pi}(\mathbf{x})$ and
$H_{ferm}(\mathbf{x})$ the fields and their conjugates by
expansions (\ref{3.49})-(\ref{3.51}) we arrive to the operators of
the no-interaction Hamiltonian
\[
\ H_{F}=H_{\pi}+H_{ferm}
\]
with
\begin{equation}
H_{\pi}=\int\omega_{\mathbf{k}}a^{\dag}(\mathbf{k})a(\mathbf{k})d\mathbf{k}\label{A.1}
\end{equation}
and
\begin{equation}
H_{ferm}=\int \!\!\!\!\!\!\!\sum
E_{\mathbf{p}}(b^{\dag}(\mathbf{p}\mu)b(\mathbf{p}\mu)+
d^{\dag}(\mathbf{p}\mu)d(\mathbf{p}\mu))d\mathbf{p}, \label{A.2}
\end{equation}
the linear momentum
$\mathbf{P}=\mathbf{P}_{F}=\mathbf{P}_{\pi}+\mathbf{P}_{ferm}$
with
\begin{equation}
\mathbf{P}_{\pi}=\int\mathbf{k}a^{\dag}(\mathbf{k})a(\mathbf{k})d\mathbf{k}
\label{A.3}
\end{equation}
and
\begin{equation}
\mathbf{P}_{ferm}=\int \!\!\!\!\!\!\!\sum
\mathbf{p}(b^{\dag}(\mathbf{p}\mu)b(\mathbf{p}\mu)+
d^{\dag}(\mathbf{p}\mu)d(\mathbf{p}\mu))d\mathbf{p}, \label{A.4}
\end{equation}
the angular momentum
$\mathbf{J}=\mathbf{J}_{F}=\mathbf{J}_{\pi}+\mathbf{J}_{ferm}$
with
\begin{equation}
\mathbf{J}_{\pi}=\frac{i}{2}\int d\mathbf{k}\mathbf{k}\times\left(
\frac{\partial a^{\dagger }(\mathbf{k})}{\partial
\mathbf{k}}a(\mathbf{k})-a^{\dagger }(\mathbf{k})\frac{\partial
a(\mathbf{k})}{\partial \mathbf{k}}\right) \label{A.5}
\end{equation}
and $\mathbf{J}_{ferm}=\mathbf{L}_{ferm}+\mathbf{S}_{ferm}$, where
\[
\ \mathbf{L}_{ferm}=\frac{i}{2}\int \!\!\!\!\!\!\!\sum
d\mathbf{p}\mathbf{p}\times\left( \frac{\partial b^{\dagger
}(\mathbf{p}\mu )}{\partial \mathbf{p}}b(\mathbf{p}\mu
)-b^{\dagger }(\mathbf{p}\mu )\frac{\partial b( \mathbf{p}\mu
)}{\partial \mathbf{p}}\right.
\]
\begin{equation}
+\left. \frac{\partial d^{\dagger }(\mathbf{p}\mu )}{\partial
\mathbf{p}}d(\mathbf{p}\mu )-d^{\dagger }(\mathbf{p}\mu
)\frac{\partial d( \mathbf{p}\mu )}{\partial
\mathbf{p}}\right),\label{A.6}
\end{equation}
\begin{equation}
\mathbf{S}_{ferm}=\frac{1}{2}\int \!\!\!\!\!\!\!\sum
d\mathbf{p}\chi^{\dag}(\mu')\mathbf{\sigma}\chi(\mu) (b^{\dagger
}(\mathbf{p}\mu')b(\mathbf{p}\mu )-d^{\dagger
}(\mathbf{p}\mu')d(\mathbf{p}\mu )),\label{A.7}
\end{equation}
the boosts $\mathbf{N}_F=\mathbf{N}_{\pi}+\mathbf{N}_{ferm}$ with
\begin{equation}
\mathbf{N}_{\pi}=\frac{i}{2}\int
d\mathbf{k}\omega_\mathbf{k}(\frac{\partial a^{\dagger
}(\mathbf{k})}{\partial
\mathbf{k}}a(\mathbf{k})-a^{\dagger}(\mathbf{k})\frac{\partial
a(\mathbf{k})}{\partial \mathbf{k}})\label{A.8}
\end{equation}
and
$\mathbf{N}_{ferm}=\mathbf{N}_{ferm}^{orb}+\mathbf{N}_{ferm}^{spin}$,
where
\[
\ \mathbf{N}_{ferm}^{orb} =\frac{i}{2}\int \!\!\!\!\!\!\!\sum
d\mathbf{p}E_\mathbf{p}\left( \frac{\partial b^{\dagger
}(\mathbf{p}\mu )}{\partial \mathbf{p}}b(\mathbf{p}\mu
)-b^{\dagger }(\mathbf{p}\mu )\frac{\partial b(\mathbf{p}\mu
)}{\partial \mathbf{p}}\right.
\]
\begin{equation}
+\left. \frac{\partial d^{\dagger }(\mathbf{p}\mu )}{\partial
\mathbf{p}}d(\mathbf{p}\mu )-d^{\dagger }(\mathbf{p}\mu
)\frac{\partial d( \mathbf{p}\mu )}{\partial
\mathbf{p}}\right),\label{A.9}
\end{equation}
\begin{equation}
\mathbf{N}_{ferm}^{spin}=-\frac 12\int \!\!\!\!\!\!\!\sum
d\mathbf{p}\mathbf{p}\times\frac{ \chi^{\dag}(\mu')\mathbf{\sigma
}\chi(\mu)}{E_{\mathbf{p}}+m}\left( b^{\dagger }( \mathbf{p}\mu'
)b(\mathbf{p}\mu )+d^{\dagger }(\mathbf{p}\mu')d(\mathbf{p} \mu
)\right). \label{A.10}
\end{equation}
In these formulae
$\omega_{\mathbf{k}}=\sqrt{\mathbf{k}^{2}+\mu_{\pi}^{2}}$
$(E_{\mathbf{p}}=\sqrt{\mathbf{p}^{2}+m^{2}})$ the pion (nucleon)
energy and $\chi(\mu)$ the Pauli spinor. When deriving Eqs.
(\ref{A.7}) and (\ref{A.10}) we have used the relations
\[
\ u^{\dagger }(\mathbf{p}\mu' )\frac{\partial u(\mathbf{p}\mu )}{
\partial \mathbf{p}}-\frac{\partial u^{\dagger }(\mathbf{p}\mu')}{\partial \mathbf{p}}u(\mathbf{
p}\mu )
\]
\begin{equation}
=v^{\dagger }(\mathbf{p}\mu' )\frac{\partial v(\mathbf{p}\mu )}{
\partial \mathbf{p}}-\frac{\partial v^{\dagger }(\mathbf{p}\mu' )}{\partial \mathbf{p}}v(\mathbf{
p}\mu )=i\frac
{\chi^{\dag}(\mu')\mathbf{\sigma}\chi(\mu)}{m(E_{\mathbf{p}}+m)}\times
\mathbf{p}.\label{A.11}
\end{equation}
with the orthonormalization conditions
\[
\ u^{\dagger }(\mathbf{p}\mu ^{\prime })u(\mathbf{p}\mu )
=v^{\dagger }(- \mathbf{p}\mu ^{\prime })v(-\mathbf{p}\mu
)=\frac{E_{\mathbf{p}}}m\delta _{\mu\mu'},
\]
\[
\ u^{\dagger }(\mathbf{p}\mu ^{\prime })v(-\mathbf{p}\mu )
=v^{\dagger }(- \mathbf{p}\mu ^{\prime })u(\mathbf{p}\mu )=0.
\]
such as in \cite{BjorkenDrell}.

Strictly speaking the fundamental relations
(\ref{2.9})-(\ref{2.11}) should be verified for every field
theory. In this connection, let us check that
\begin{equation}
\ [P^{j},N_{F}^{l}]=i\delta_{jl}H_{F}.\label{A.12}
\end{equation}
In fact, we find step by step
\[
\
[P^{j},N_{F}^{l}]=[P^{j}_{\pi},N_{\pi}^{l}]+[P^{j}_{ferm},N_{ferm}^{l}],
\]
\[
\ [P^{j}_{\pi},N_{\pi}^{l}]= -i\frac{\partial}{\partial
u^{j}}\left\{e^{i\mathbf{P}_{\pi}\mathbf{u}}N_{\pi}^{l}e^{-i\mathbf{P}_{\pi}\mathbf{u}}
\right\}|_{\mathbf{u}=0}
\]
\[
=\frac{1}{2}\frac{\partial}{\partial u^{j}}\int
d\mathbf{k}\omega_\mathbf{k}\left(\frac{\partial}{\partial
k^{l}}[e^{i\mathbf{u}\mathbf{k}}a^{\dagger
}(\mathbf{k})]a(\mathbf{k})e^{-i\mathbf{u}\mathbf{k}}-e^{i\mathbf{u}\mathbf{k}}a^{\dagger
}(\mathbf{k})\frac{\partial}{\partial
k^{l}}[e^{-i\mathbf{u}\mathbf{k}}a(\mathbf{k})]\right)|_{\mathbf{u}=0}
\]
\[
 =\frac{\partial}{\partial u^{j}}\left\{i\int \omega_{\mathbf{k}}d\mathbf{k}u^{l}a^{\dag}(\mathbf{k})a(\mathbf{k})-i\mathbf{N}_{\pi}^{l}\right\}
_{\mathbf{u}=0}=i\delta_{jl}H_{\pi},
\]
\[
[P^{j}_{ferm},N_{ferm}^{l}]=[P^{j}_{ferm},N_{ferm}^{orb,l}]
\]
\[
 =-\frac{\partial}{\partial u^{j}}\int \!\!\!\!\!\!\!\sum d\mathbf{p}E_\mathbf{p}(\frac{\partial}{\partial
p^{l}}[e^{i\mathbf{u}\mathbf{p}}b^{\dagger
}(\mathbf{p}\mu)]b(\mathbf{p}\mu)e^{-i\mathbf{u}\mathbf{p}}-e^{i\mathbf{u}\mathbf{p}}b^{\dagger
}(\mathbf{p}\mu)\frac{\partial}{\partial
p^{l}}[e^{-i\mathbf{u}\mathbf{p}}b(\mathbf{p}\mu)]
\]
\[
 +b^{\dagger }(\mathbf{p}\mu)\rightarrow d^{\dagger }(\mathbf{p}\mu),b(\mathbf{p}\mu)\rightarrow d(\mathbf{p}\mu))|_{\mathbf{u}=0}
\]
\[
 =\frac{\partial}{\partial u^{j}}\left\{iu^{l}\int \!\!\!\!\!\!\!\sum E_{\mathbf{p}}d\mathbf{p}(b^{\dag}(\mathbf{p}\mu)b(\mathbf{p}\mu)+d^{\dag}(\mathbf{p}\mu)d(\mathbf{p}\mu))-i\mathbf{N}_{ferm}^{l}\right\}
_{\mathbf{u}=0}=i\delta_{jl}H_{ferm}.
\]
Analogously,
\begin{equation}
\
[H_{F},\mathbf{N}_{F}]=-i\frac{d}{d\lambda}\left\{e^{iH_{F}\lambda}N_{F}e^{-iH_{F}\lambda}
\right\}|_{\lambda=0}=i\mathbf{P}.\label{A.13}
\end{equation}
We also need the expression
\begin{equation}
\mathbf{N}_{mes}=\frac{i}{2}\int
d\mathbf{k}'d\mathbf{k}a^{\dagger}(\mathbf{k}')a(\mathbf{k})
\frac{\omega_\mathbf{k'}\omega_\mathbf{k}+\mathbf{k}'\mathbf{k}+\mu_{s}^{2}}
{\sqrt{\omega_\mathbf{k'}\omega_\mathbf{k}}}\frac{\partial}{\partial\mathbf{k}}\delta(\mathbf{k}-\mathbf{k}'),
\label{A.14}
\end{equation}
equivalent to (\ref{A.8}) and the free boost
\begin{equation}
\mathbf{N}_{bos}=\frac{i}{2}\int
d\mathbf{p}'d\mathbf{p}(b^{\dagger}(\mathbf{p}')b(\mathbf{p})+d^{\dagger}(\mathbf{p}')d(\mathbf{p}))
\frac{E_\mathbf{p'}E_\mathbf{p}+\mathbf{p}'\mathbf{p}+\mu_{b}^{2}}
{\sqrt{E_\mathbf{p'}E_\mathbf{p}}}\frac{\partial}{\partial\mathbf{p}}\delta(\mathbf{p}-\mathbf{p}'),
\label{A.15}
\end{equation}
for the spinless charged bosons.

\section{Evaluation of commutator $[V_{\pi N}(\mathbf{x}'),V_{\pi N}(\mathbf{x})]$ with a nonlocal $\pi N$ interaction}

Let us rewrite the $\pi N$ interaction density in the r.h.s of Eq.
(\ref{3.47}) as
\begin{equation}
V_{ps}(\mathbf{x})\equiv
V_{loc}(\mathbf{x})=\varphi_{ps}(\mathbf{x})f_{loc}(\mathbf{x}),
\label{B.1}
\end{equation}
\begin{equation}
f_{loc}(\mathbf{x})=ig\frac{m}{(2\pi)^{3}}\int\frac{d\mathbf{p}'}{E_{\mathbf{p}'}}\int\frac{d\mathbf{p}}{E_{\mathbf{p}}}
[\bar{B}(p'),\bar{D}(p')]\gamma_{5} \left[
\begin{array}{l}
B(p) \\
D(p)
\end{array}
\right] e^{-i(\mathbf{p'}-\mathbf{p})\mathbf{x}} \label{B.2}
\end{equation}
with notations
\[
B(p)=\sum_{\mu}u(p\mu)b(p\mu),
\]
\[
D(p)=\sum_{\mu}v(p_{-}\mu)d^{\dag}(p_{-}\mu)
\]
and commutations
\begin{equation}
\{B_{a}(p'),\bar{B}_{b}(p)\}=p_{0}\delta(\mathbf{p'}-\mathbf{p})(a|P_{+}(p)|b),
\label{B.3}
\end{equation}
\begin{equation}
\{D_{a}(p'),\bar{D}_{b}(p)\}=p_{0}\delta(\mathbf{p'}-\mathbf{p})(a|P_{-}(p_{-})|b),
\label{B.4}
\end{equation}
where $a$ and $b$ spinor indices and
\[
P_{\pm}(p)=\frac{\hat{p}\pm m}{2m}
\]
the standard projection operators.

Here we will consider a nonlocal extension of the Yukawa-type
density (\ref{B.1}) by introducing \footnote{Henceforth, such an
occurrence in $V_{nloc}(\mathbf{x})$ of the "coordinate"
$\mathbf{x}$ and the subscript {\it nloc} does not contradict each
other. The former originates from translational invariance (cf.
the transition from Eq. (\ref{1.6}) to Eq. (\ref{2.14})) while the
latter allows us to work with the interaction density not being
constructed from fields (in our case $\bar{\psi}$ and $\psi$)
which are taken at one and the same point. A similar nonlocal
interaction one can find in \cite{Shir2002} (see Eq. (4.45)
therein).}
\begin{equation}
V_{nloc}(\mathbf{x})=\varphi_{ps}(\mathbf{x})f_{nloc}(\mathbf{x}),
\label{B.5}
\end{equation}
\begin{equation}
f_{nloc}(\mathbf{x})=i\frac{m}{(2\pi)^{3}}\int\frac{d\mathbf{p}'}{E_{\mathbf{p}'}}\int\frac{d\mathbf{p}}{E_{\mathbf{p}}}
g(p',p)[\bar{B}(p'),\bar{D}(p')]\gamma_{5} \left[
\begin{array}{l}
B(p) \\
D(p)
\end{array}
\right] e^{-i(\mathbf{p'}-\mathbf{p})\mathbf{x}} \label{B.6}
\end{equation}
with a real and permutably symmetric cutoff function $g(p',p)$.
One more condition,
\begin{equation}
g(\Lambda p',\Lambda p)=g(p',p) \label{B.7}
\end{equation}
allows for the operator $V_{nloc}(x)$ to be the Lorentz scalar.
Since
\[
[V_{nloc}(\mathbf{x}),V_{nloc}(\mathbf{y})]=\varphi(\mathbf{x})\varphi(\mathbf{y})
[f_{nloc}(\mathbf{x}),f_{nloc}(\mathbf{y})],
\]
the requirement in question
\begin{equation}
[V_{nloc}(\mathbf{x}),V_{nloc}(\mathbf{y})]=0 \label{B.8}
\end{equation}
is equivalent to
\begin{equation}
[f_{nloc}(\mathbf{x}),f_{nloc}(\mathbf{y})]=0. \label{B.9}
\end{equation}
At this point, using that technique from Appendix A of
\cite{SheShi01} with the aid of the identities
\[
[AB,CD]=A\{B,C\}D-\{A,C\}BD-C\{D,A\}B+CA\{D,B\},
\]
\begin{equation}
[AB,CD]=A[B,C]D+[A,C]DB+AC[B,D]+C[A,D]B. \label{B.10}
\end{equation}
for four operators $A$, $B$, $C$ and $D$, one can show that
\begin{equation}
[f_{nloc}(\mathbf{x}),f_{nloc}(\mathbf{y})]=
\frac{m}{(2\pi)^{3}}\int\frac{d\mathbf{p}'}{E_{\mathbf{p}'}}\int\frac{d\mathbf{p}}{E_{\mathbf{p}}}
e^{-i\mathbf{p'}\mathbf{x}+i\mathbf{p}\mathbf{y}}f_{nloc}(\mathbf{x}-\mathbf{y};p',p)-H.c.
\label{B.11}
\end{equation}
with
\[
f_{nloc}(\mathbf{x}-\mathbf{y};p',p)=-\frac{m}{(2\pi)^{3}}\int\frac{d\mathbf{q}}{E_{\mathbf{q}}}
e^{i\mathbf{q}(\mathbf{x}-\mathbf{y})}
\]
\[
\times g(p',q)g(p,q)[\bar{B}(p')+\bar{D}(p')]\gamma_{5}
[P_{+}(q_{+})+P_{-}(q_{-})]\gamma_{5}[B(p)+D(p)]
\]
or
\begin{equation}
f_{nloc}(\mathbf{x}-\mathbf{y};p',p)=g(\mathbf{x}-\mathbf{y};p',p)[B^{\dag}(p')+D^{\dag}(p')][B(p)+D(p)],
\label{B.12}
\end{equation}
where
\begin{equation}
g(\mathbf{x}-\mathbf{y};p',p)=\frac{1}{(2\pi)^{3}}\int
d\mathbf{q}e^{i\mathbf{q}(\mathbf{x}-\mathbf{y})}g(p',q)g(p,q).
\label{B.13}
\end{equation}
Putting $g(p',p)\equiv g$ that yields
$g(\mathbf{z};p',p)=\delta(\mathbf{z})$, we come back to the
initial local model with its property
\begin{equation}
[f_{loc}(\mathbf{x}),f_{loc}(\mathbf{y})]=0. \label{B.14}
\end{equation}
In order to go out beyond the model one can regard the two
options,
\begin{equation}
g(p',p)=gexp[\frac{(p'-p)^{2}}{2\Lambda^{2}}]=g(\Lambda)exp[-\frac{p'p}{\Lambda^{2}}]
\label{B.15}
\end{equation}
and
\begin{equation}
g(p',p)=g\frac{\Lambda^{2}-\mu^{2}_{\pi}}{\Lambda^{2}+(p'-p)^{2}}.
\label{B.16}
\end{equation}
Here we will restrict ourselves to the first using the second for
other applications. In the context, it is convenient to deal with
the invariants
\begin{equation}
I^{(\pm)}(z;p',p)=\int\frac{d\mathbf{q}}{E_{\mathbf{q}}}e^{\mp
iqz}g(p',q)g(p,q)=I^{(\pm)}(\Lambda z;\Lambda p',\Lambda p).
\label{B.17}
\end{equation}
In case of the factor (\ref{B.15}) we encounter the integrals
\begin{equation}
I^{(\pm)}(x'-x;p',p)=g^{2}(\Lambda)\int\frac{d\mathbf{q}}{E_{\mathbf{q}}}e^{\mp
iq(x'-x)}e^{-\lambda uq}, \label{B.18}
\end{equation}
where $u=p'+p$, $\lambda = \Lambda^{-2}$ and
$g(\Lambda)=gexp(\lambda m^{2})$. Thus, since $I^{(-)*}=I^{(+)}$,
our task is to evaluate
\begin{equation}
\Delta^{(+)}(x'-x+i\lambda
u;m)=\int\frac{d\mathbf{q}}{E_{\mathbf{q}}}exp[i(x'-x+i\lambda
u)q]. \label{B.19}
\end{equation}
But from $\Delta^{(+)}(\Lambda z;m)=\Delta^{(+)}(z;m)$ it follows
that
\begin{equation}
\Delta^{(+)}(x'-x+i\lambda u;m)=\Delta^{(+)}(v+i\lambda r;m),
\label{B.20}
\end{equation}
where the Lorentz transformation $L=L(\mathbf{u})$ is such that
$Lu=(r_{0},\mathbf{0})$ with $r_{0}>0$. Recall that
$u^{2}=(p'+p)^{2}=2m^{2}+2p'p>0$. In turn, $v=L(x'-x)$.

Furthermore, it is well known (see, e.g., formula (3.961.1) in
\cite{GR}) that
\[
\int_{0}^{\infty}e^{-\beta\sqrt{\gamma^{2}+y^{2}}}\sin{ay}\frac{ydy}{\sqrt{\gamma^{2}+y^{2}}}=
\frac{ay}{\sqrt{\gamma^{2}+a^{2}}}K_{1}(\gamma\sqrt{a^{2}+\beta^{2}}),
\]
\[
[Re\beta>0, Re\gamma>0, a>0]
\]
where $K_{1}(z)$ the modified Bessel function.

Using the result we find
\[
\Delta^{(+)}(v+i\lambda
r;m)=\int\frac{d\mathbf{q}}{E_{\mathbf{q}}}e^{-E_{\mathbf{q}}(\lambda
r_{0}-iv_{0})}e^{i\mathbf{q}\mathbf{v}}
\]
\[
=4\pi\int_{0}^{\infty}\frac{qdq}{\sqrt{q^{2}+m^{2}}}
e^{-\sqrt{q^{2}+m^{2}}(\lambda
y_{0}-iv_{0})}\frac{\sin(q|\mathbf{v}|)}{|\mathbf{v}|}
\]
\begin{equation}
= 4\pi m^{2}\frac{K_{1}(z_{0})}{z_{0}} \label{B.21}
\end{equation}
with $z_{0}=m\sqrt{\lambda^{2}u^{2}-(x'-x)^{2}-2i\lambda
v_{0}\sqrt{u^{2}}}$. In the case of interest
$x'-x=(0,\mathbf{x}-\mathbf{y})$ and
$v_{0}=L^{0}_{j}(\mathbf{x}-\mathbf{y})^{j}=\frac{\mathbf{u}(\mathbf{x}-\mathbf{y})}{\sqrt{u^{2}}}$,
so
$z_{0}=m\sqrt{\lambda^{2}u^{2}+(\mathbf{x}-\mathbf{y})^{2}-2i\lambda
\mathbf{u}(\mathbf{x}-\mathbf{y})}=\varsigma_{0}$ and
\begin{equation}
g(\mathbf{x}-\mathbf{y};p',p)=4\pi i m^{4}\lambda
u^{0}\frac{K_{2}(\varsigma_{0})}{\varsigma_{0}^{2}}. \label{B.22}
\end{equation}
Here we have employed the matrix
\[
\ L^{\mu}_{\,\,\nu}(u)= \left[
\begin{array}{ll}
\frac{u^0}{\sqrt{u^{2}}} & \frac{u^j}{\sqrt{u^{2}}} \\
-\frac{u^i}{\sqrt{u^{2}}} & \delta
_j^i-\frac{u^iu_j}{u^{2}+\sqrt{u^{2}}u^{0}}
\end{array}
\right].
\]
Formula (\ref{B.22}) suffices for the statement below
(\ref{3.70}).

\section{Evaluation of integral $m_{s}^{(2)}(k)$}
The alternative in question is prompted by Pauli and Rose
\cite{PauliRose} with their refined trick to be applied to
\begin{equation}
m_{s}^{(2)}(k)=\int
d\mathbf{p}\frac{E_{\mathbf{p}+\frac{\mathbf{k}}{2}}+
E_{\mathbf{p}-\frac{\mathbf{k}}{2}}}{E_{\mathbf{p}+\frac{\mathbf{k}}{2}}E_{\mathbf{p}-\frac{\mathbf{k}}{2}}}
\frac{f^{2}(\omega_{\mathbf{k}}^{2}-(E_{\mathbf{p}+\frac{\mathbf{k}}{2}}+E_{\mathbf{p}-\frac{\mathbf{k}}{2}})^{2})}
{(E_{\mathbf{p}+\frac{\mathbf{k}}{2}}+E_{\mathbf{p}-\frac{\mathbf{k}}{2}})^{2}-\omega^{2}_{\mathbf{k}}}.
\label{C.1}
\end{equation}

In order to go on let us introduce the new variables $w$, $v$ and
$\varphi$, where $\varphi$ is the azimuthal angle around the axis
parallel to $\mathbf{k}$, so
\begin{equation}
\frac{1}{2}(E_{\mathbf{p}+\frac{\mathbf{k}}{2}}+E_{\mathbf{p}-\frac{\mathbf{k}}{2}})=
w,\,\,\,\,\,\,
\frac{1}{2}(E_{\mathbf{p}+\frac{\mathbf{k}}{2}}-E_{\mathbf{p}-\frac{\mathbf{k}}{2}})=
v. \label{C.2}
\end{equation}
Using the corresponding Jacobian, we obtain
\begin{equation}
\frac{1}{4}\frac{E_{\mathbf{p}+\frac{\mathbf{k}}{2}}+
E_{\mathbf{p}-\frac{\mathbf{k}}{2}}}{E_{\mathbf{p}+\frac{\mathbf{k}}{2}}E_{\mathbf{p}-\frac{\mathbf{k}}{2}}}
d\mathbf{p}= \frac{w}{k}dvdwd\varphi. \label{C.3}
\end{equation}
From (\ref{C.2}) we get
\begin{equation}
\frac{1}{2}(E_{\mathbf{p}+\frac{\mathbf{k}}{2}}^{2}+E_{\mathbf{p}-\frac{\mathbf{k}}{2}}^{2})=w^{2}+v^{2}=
\mathbf{p}^{2}+\frac{\mathbf{k}^{2}}{4}+\mu_{b}^{2}, \label{C.4}
\end{equation}
\begin{equation}
\frac{1}{4}(E_{\mathbf{p}+\frac{\mathbf{k}}{2}}^{2}-E_{\mathbf{p}-\frac{\mathbf{k}}{2}}^{2})=wv=
\frac{1}{2}pk\cos\varphi. \label{C.5}
\end{equation}
The limits of integration over the new variables are
\[
0\leq\varphi\leq2\pi,
\]
\[
v_{-}=-\frac{k}{2}\sqrt{\frac{w^{2}-\frac{k^{2}}{4}-\mu_{b}^{2}}{w^{2}-\frac{k^{2}}{4}}}\leq
v\leq
v_{+}=\frac{k}{2}\sqrt{\frac{w^{2}-\frac{k^{2}}{4}-\mu_{b}^{2}}{w^{2}-\frac{k^{2}}{4}}},
\]
\begin{equation}
w_{0}\equiv\sqrt{\frac{k^{2}}{4}+\mu_{b}^{2}}\leq w\leq\infty.
\label{C.6}
\end{equation}
By integrating in (\ref{C.1}) we arrive to
\[
m_{s}^{(2)}(k)=4\int_{w_{0}}^{\infty}dw\int_{v_{-}}^{v_{+}}
dv\int_{0}^{2\pi}d\varphi\frac{1}{k}\frac{w}{4w^{2}-\omega^{2}_{\mathbf{k}}}
f^{2}(\omega^{2}_{\mathbf{k}}-4w^{2})
\]
\[
=8\pi\int_{w_{0}}^{\infty}\sqrt{\frac{w^{2}-\frac{k^{2}}{4}-\mu_{b}^{2}}{w^{2}-\frac{k^{2}}{4}}}
\frac{wdw}{4w^{2}-\omega^{2}_{\mathbf{k}}}f^{2}(\omega^{2}_{\mathbf{k}}-4w^{2})
\]
\[
=8\pi\int_{\mu_{b}}^{\infty}d\epsilon\frac{\sqrt{\epsilon^{2}-\mu_{b}^{2}}}{4\epsilon^{2}-\mu_{s}^{2}}
f^{2}(\mu_{s}^{2}-4\epsilon^{2})
\]
\begin{equation}
=8\pi\int_{0}^{\infty}\frac{t^{2}dt}{\sqrt{t^{2}+\mu_{b}^{2}}}\frac{f^{2}(\mu_{s}^{2}-4t^{2}-4\mu_{b}^{2})}{4t^{2}+4\mu_{b}^{2}-\mu_{s}^{2}}
\label{C.7}
\end{equation}
that coincides with the formula (\ref{5.200}). Using the popular
form (\ref{B.16}) we have the cutoff function
\begin{equation}
g_{12}(p,q_{-},k)=g\frac{\Lambda^{2}-\mu_{s}^{2}}{\Lambda^{2}-(p-q_{-})^{2}}
\label{C.8}
\end{equation}
that in combination with assumption (\ref{5.198}) is equivalent to
the relation
\begin{equation}
f(I)=g\frac{\Lambda^{2}-\mu_{s}^{2}}{\Lambda^{2}+\mu_{s}^{2}-4\mu_{b}^{2}-I}
 \label{C.9}
\end{equation}
and gives the expression
\[
 m_{s}^{(2)}=2\pi g^{2}\frac{(\Lambda^{2}-\mu_{s}^{2})^{2}}{\Lambda^{4}}
\]
\[
 \times \left[
\frac{\Lambda^{2}(4\mu_{b}^{2}-\mu_{s}^{2})}{(\Lambda^{2}-4\mu_{b}^{2}+\mu_{s}^{2})^{2}}
\left(\frac{\Lambda}{\sqrt{4\mu_{b}^{2}-\Lambda^{2}}}
\arctan\frac{\sqrt{4\mu_{b}^{2}-\Lambda^{2}}}{\Lambda}
-\frac{\Lambda^{2}}{\mu_{s}\sqrt{4\mu_{b}^{2}-\mu_{s}^{2}}}
\arctan\frac{\mu_{s}}{\sqrt{4\mu_{b}^{2}-\mu_{s}^{2}}}\right)
\right.
\]
\begin{equation}
 +\frac{\Lambda^{2}}{\Lambda^{2}-4\mu_{b}^{2}+\mu_{s}^{2}} \left(
\frac{\Lambda(2\mu_{b}^{2}-\Lambda^{2})}{(4\mu_{b}^{2}-\Lambda^{2})^{3/2}}
\arctan\frac{\sqrt{4\mu_{b}^{2}-\Lambda^{2}}}{\Lambda} \right.
\left.\left.+\frac{1}{2}\frac{\Lambda^{2}}{4\mu_{b}^{2}-\Lambda^{2}}\right)\right],
\label{C.10}
\end{equation}
if $\Lambda<2\mu_{b}$

and
\[
 m_{s}^{(2)}=2\pi g^{2}\frac{(\Lambda^{2}-\mu_{s}^{2})^{2}}{\Lambda^{4}}
\]
\[
 \times \left[
\frac{\Lambda^{2}(4\mu_{b}^{2}-\mu_{s}^{2})}{(\Lambda^{2}-4\mu_{b}^{2}+\mu_{s}^{2})^{2}}
\left(\frac{\Lambda}{2\sqrt{\Lambda^{2}-4\mu_{b}^{2}}}
\ln\frac{\Lambda+\sqrt{\Lambda^{2}-4\mu_{b}^{2}}}
{\Lambda-\sqrt{\Lambda^{2}-4\mu_{b}^{2}}}
-\frac{\Lambda^{2}}{\mu_{s}\sqrt{4\mu_{b}^{2}-\mu_{s}^{2}}}
\arctan\frac{\mu_{s}}{\sqrt{4\mu_{b}^{2}-\mu_{s}^{2}}}\right)
\right.
\]
\begin{equation}
 +\frac{\Lambda^{2}}{\Lambda^{2}-4\mu_{b}^{2}+\mu_{s}^{2}} \left(
\frac{\Lambda(\Lambda^{2}-2\mu_{b}^{2})}{2(\Lambda^{2}-4\mu_{b}^{2})^{3/2}}
\ln\frac{\Lambda+\sqrt{\Lambda^{2}-4\mu_{b}^{2}}}
{\Lambda-\sqrt{\Lambda^{2}-4\mu_{b}^{2}}}
 \right.
\left.\left.+\frac{1}{2}\frac{\Lambda^{2}}{4\mu_{b}^{2}-\Lambda^{2}}\right)\right],
\label{C.11}
\end{equation}
if $\Lambda>2\mu_{b}$.

By putting $\mu_{s}=\mu_{\pi}=0.6994 fm^{-1}$ and
$\mu_{b}=m=4.7583 fm^{-1}$ we find the following sequence
\[
m_{s}^{(2)}10^{4}/2\pi
g^{2}=1.853,\,\,\,34.69,\,\,\,109.84,\,\,\,224.74,\,\,\,335.05,...
\]
at $\Lambda=1,\,\,2,\,\,3,\,\,4,\,\,\mu_{b},...$.
 All values in $fm^{-1}$.

\end{document}